\documentclass[12pt]{article}

\usepackage{amssymb}
\usepackage{latexsym}

\def \be{\begin{equation}}
\def \ee{\end{equation}}
\def \berr{\begin{eqnarray}}
\def \err{\end{eqnarray}}
\def \nn{\nonumber}
\def \a{\alpha}

\def \b{\beta}
\def \g{\gamma}
\def \d{\delta}

\def \Del{\Delta}

\def \eps{\varepsilon}

\def \l{\lambda}
\def \tp{\tilde\phi}

\def \hf{\frac 12}

\def \dl{\partial}
\def \p{\varphi}

\def \A{{\cal A}}
\def \D{{\cal D}}
\def \F{{\cal F}}

\def \H{{\cal H}}

\def \S{{\cal S}^2}

\def \RR{{\cal R}}
\def \TR{\tilde{\cal R}}
\def \NN{{\cal N}}

\def \({\left(}
\def \){\right)}
\def \<{\langle}
\def \>{\rangle}
\def \[{\left[}
\def \]{\right]}

\def \obar{\overline}

\def \wtilde{\widetilde}
\def \del{\partial}
\def\tens{\mathop{\otimes}}

\def\id{\rm id}
\newcommand{\tr}{\triangleright}
\newcommand{\ttr}{\tilde\triangleright}

\newcommand \one{{\bf 1}}

\def\reps{representations }
\def\rep{representation }

\def\mf#1{{\mathbb #1}}

\def\R{{\mf{R}}}
\def\C{{\mf{C}}}
\newcommand \reals{\R}

\def\smash{\mbox{$\,\rule{0.3pt}{1.1ex}\!\times\,$}}

\newcommand{\sect}[1]{\setcounter{equation}{0}\section{#1}}

\newtheorem{prop}{Proposition}[section]

\newtheorem{lemma}[prop]{Lemma}

\evensidemargin \oddsidemargin
\begin{document}
\begin{titlepage}
\begin{flushright}
LMU-TPW 1-01 \\
LPT-ORSAY 01-22\\
UWThPh-2001-10

\end{flushright}
\begin{center}
{\Large \bf Field Theory\\ on the  $q$--deformed Fuzzy Sphere II: \\[1ex]
           Quantization}\\[20pt]
H. Grosse$^a$\footnote{grosse@doppler.thp.univie.ac.at}, 
   J.\  Madore$^{b,c}$\footnote{John.Madore@th.u-psud.fr}  and  
H.\ Steinacker$^{c,d}$\footnote{Harold.Steinacker@th.u-psud.fr} \\[2ex] 
{\small\it 
        ${}^a$Institut for Theoretical Physics, University of Vienna,\\
              Boltzmanngasse 5, A-1090 Vienna, Austria \\[2ex]

        ${}^b$Max-Planck-Institut f\"ur Physik\\
           F\"ohringer Ring 6, D-80805 M\"unchen  \\[2ex]
        
        ${}^c$Laboratoire de Physique Th\'eorique et Hautes Energies\\
        Universit\'e de Paris-Sud, B\^atiment 211, F-91405 Orsay \\[2ex]

       ${}^d$Sektion Physik der Ludwig--Maximilians--Universit\"at M\"unchen\\
        Theresienstr.\ 37, D-80333 M\"unchen  \\[1ex] }

{\bf Abstract} \\
\end{center}

We study the second quantization of field theory on the 
$q$--deformed fuzzy sphere for $q \in \R$. This is performed using a 
path integral over the modes, which generate a quasiassociative algebra.
The resulting models have a manifest $U_q(su(2))$ symmetry with 
a smooth limit $q \rightarrow 1$, and satisfy positivity and twisted bosonic 
symmetry properties. A systematic way to calculate $n$--point correlators 
in perturbation theory is given. As examples, the 
4--point correlator for a free scalar field theory and the planar 
contribution to the tadpole diagram in $\phi^4$ theory are computed. 
The case of gauge fields is also discussed, as well as an operator 
formulation of scalar field theory in $2_q + 1$ dimensions. An alternative, 
essentially equivalent approach using associative techniques only is also 
presented. The proposed framework is not restricted to 2 dimensions.

\noindent
 
 

\end{titlepage}

\tableofcontents

\sect{Introduction}

The idea of studying field theory on $q$--deformed spaces has been 
pursued since their 
appearance more than 10 years ago. While much work has
been done on the level of first quantization (see for example 
\cite{bianca,fiore,ours_1,weich,majid_braiding,meyer,podles_wave} 
and references therein),
the second quantization has proved to be difficult. The main 
problem is perhaps the apparent incompatibility between the 
symmetrization postulate of quantum field theory (QFT), 
and the fact that quantum groups are naturally associated with the
braid group rather than the symmetric group. One could of course 
consider theories with generalized statistics; however if 
$q$--deformation is considered as a true ``deformation'' of ordinary space,
then it should be possible to define models with a smooth limit 
$q \rightarrow 1$. In particular, the number of degrees of freedom 
should be independent of $q$. The goal of the present paper is to define 
a $q$--deformed QFT which is essentially bosonic, and has 
a smooth limit $q \rightarrow 1$.

In our previous work \cite{ours_1}, we studied in detail 
field theory on the $q$--deformed fuzzy spheres $\S_{q,N}$ 
at the first--quantized level. The spheres $\S_{q,N}$ are
precisely the ``discrete series'' of Podles spheres \cite{podles} 
if $q \in \R$.
This space is particularly well suited to attack the problem of
second quantization, because there is only a finite number of modes.
Therefore all considerations can be done on a purely algebraic 
level, and are essentially rigorous.
The methods we shall develop here are however not restricted to that case, 
but should generalize immediately
to other $q$--deformed spaces, at least on a formal level. 
There will be complications, of course, if the number of 
modes is infinite. 

To understand the problem, consider scalar fields,
which are elements $\psi \in \S_{q,N}$. 
A typical action can have the form
\be
S[\psi] = - \int\limits_{\S_{q,N}}\hf\;\psi^{\ast}\Delta \psi + \lambda\psi^4,
\ee
where $\Delta$ is the Laplacian \cite{ours_1}. Such actions are
invariant under the quantum group $U_q(su(2))$ of rotations, and they are
real, $S[\psi]^* = S[\psi]$. 
They define a first--quantized euclidean scalar
field theory on the $q$--deformed fuzzy sphere. 

We want to study the second quantization of these models. 
On the undeformed fuzzy sphere, this is fairly straightforward 
\cite{grosse,madore_sphere}:
The fields can be expanded in terms of irreducible \reps of $SO(3)$,
\be
\psi(x) = \sum_{K,n} \psi_{K,n}(x) \;  a^{K,n} 
\label{psi_field_intro}
\ee
with coefficients $a^{K,n} \in \C$. The above actions then become 
polynomials in the variables $a^{K,n}$
which are invariant under $SO(3)$, 
and the ``path integral'' is naturally defined as 
the product of the ordinary integrals over the coefficients $a^{K,n}$.
This defines a quantum field theory which has a $SO(3)$ rotation symmetry, 
because the path integral is invariant.

In the $q$--deformed case, this is not so easy.
The reason is that the coefficients $a^{K,n}$ in (\ref{psi_field_intro}) 
must be considered as \reps of $U_q(su(2))$ in order 
to have such a symmetry at the quantum level. This implies that
they cannot be ordinary complex numbers, because
a commutative algebra is not consistent with the action of $U_q(su(2))$,
whose coproduct is not cocommutative. Therefore an ordinary
integral over commutative modes $a^{K,n}$ would violate $U_q(su(2))$ 
invariance at the quantum level. On the other hand, no associative 
algebra with generators $a^{K,n}$ is known (except for some simple 
representations) which is both covariant under 
$U_q(su(2))$ and has the same Poincar\'e series as classically, 
i.e. the dimension of the space of polynomials at a given degree is 
the same as in the undeformed case. The 
latter is an essential physical requirement at least for 
low energies, in order to have the correct number of degrees of freedom,
and is usually encoded in a symmetrization postulate. It means that 
the ``amount of information'' contained in the $n$--point functions 
should be the same as for $q=1$. These issues will 
be discussed on a more formal level in Section \ref{sec:QFT}.
While some proposals have been given in the literature
\cite{oeckl,chaichian} how to define 
QFT on spaces with quantum group symmetry, none of them seems to
satisfies all these requirements.

One possible way out has been suggested in \cite{fiore_schupp}, where it
was shown that a symmetrization can be achieved using a  
Drinfel'd twist, at least in any given $n$--particle sector.
Roughly speaking, the Drinfel'd twist relates the tensor product of \reps
of quantum groups to the tensor product of undeformed ones, 
and hence essentially allows to use the usual completely
symmetric Hilbert space. The problem remained, however, how to 
treat sectors with different particle number simultaneously,
which is essential for a QFT, and how to handle
the Drinfel'd twists which are very difficult to calculate.

We present here a formalism which solves these problems, by
defining a star product of the modes $a^{K,n}$ which is 
covariant under the quantum group, and in the limit
$q \rightarrow 1$ reduces to the commutative algebra of functions in the
$a^{K,n}$. This algebra is quasiassociative, but satisfies all
the requirements discussed above. In particular, the number of 
independent polynomials in the $a^{K,n}$ is the same as usual.
One can then define an invariant path integral, which yields a 
consistent and physically reasonable definition of a second--quantized 
field theory with a quantum group symmetry. In particular, the 
``correlation functions'' will satisfy invariance, hermiticity, positivity
and symmetry properties. An essentially equivalent formulation in terms 
of a slightly extended associative algebra will be presented
as well, based on constructions by Fiore \cite{fiore_twist}. It turns out 
to be related to the general considerations in \cite{mack_schom}.
The appearance of quasiassociative algebras is also consistent with 
results in the context of $D$--branes on WZW models \cite{ars}.

Our considerations are not restricted to 2 dimensions, 
and should be applicable to other spaces with quantum group symmetry as well.
The necessary mathematical tools will be developed in 
Sections \ref{sec:drinfeld_twist}, \ref{sec:twisting}, and 
\ref{sec:operator-form}. After discussing the definition and basic
properties of QFT on $\S_{q,N}$ in Section \ref{sec:QFT}, we
derive formulas to calculate
$n$--point functions in perturbation theory, and find
an analog of Wick's theorem. All diagrams on $\S_{q,N}$ are of course finite,
and vacuum diagrams turn out to cancel a usual.
The resulting models can also be interpreted as field theories
on the undeformed fuzzy sphere, with slightly ``nonlocal'' interactions.

As applications of the general method, we consider first the
case of a free scalar field theory, and calculate the 4--point functions.
The tadpole diagram for a $\phi^4$ model is studied as well, and
turns out to be linearly divergent as $N \rightarrow \infty$. 
We then discuss two possible quantizations of gauge models,
and finally consider scalar field theory on $\S_{q,N}$ with an extra 
time.

We should stress that our approach is 
quite conservative, as it aims to find a ``deformation'' of standard 
quantum field theory in a rather strict sense, with ordinary statistics.
Of course on can imagine other, less conventional approaches,
such as the one in \cite{oeckl}.
Moreover, we only consider the case $q \in \R$ in this paper. It should be 
possible to modify our methods so that the 
case of $q$ being a root of unity can also be covered. 
Then QFT on more realistic spaces such as
4--dimensional quantum Anti--de Sitter space \cite{ads_space} could 
be considered as well. There, the number of modes as well as the
dimensions of the relevant \reps are finite at roots of unity, 
as in the present paper.

\sect{Some mathematical background: Drinfel'd twists}
\label{sec:drinfeld_twist}

We first review some mathematical results which are the basis of 
the later considerations. In order to avoid confusions, the language
will be quite formal initially. 
To a given a finite--dimensional simple Lie 
algebra $g$ (for our purpose just $su(2)$), one can associate 2 Hopf algebras:
the usual $(U(g)[[h]],m,\eps,\Delta,S)$,
and the $q$--deformed $(U_q(g),m_q,\eps_q,\Delta_q,S_q)$. Here
$U(g)$ is the universal enveloping algebra, 
$U_q(g)$ is the $q$--deformed universal enveloping algebra, and
$U(g)[[h]]$ are the formal power series in $h$ with 
coefficients in $U(g)$. The symbol
$$
q=e^h
$$
is considered formal for now.
Then a well-known theorem by Drinfel'd  (Proposition 3.16
in Ref. \cite{drinfeld_quasi}) states that there exists an algebra 
isomorphism
\be
\varphi: U_q(g) \rightarrow U(g)[[h]]
\label{phi}
\ee
and a `twist', i.e. an element 
$$
\F = \F_1\tens \F_2 \; \in \; U(g)[[h]]\tens U(g)[[h]]
$$
(in a Sweedler notation, where a sum is implicitly understood) 
satisfying
\berr
&&(\varepsilon\tens \id)\F=\one=(\id\tens \varepsilon)\F,
\label{cond2}\\
&&\F = \one\tens\one +o(h),
\label{cond2bis}
\err
which relates these two Hopf algebra $U_q(g)$ and $U(g)[[h]]$ 
as follows: if  $\F^{-1}=\F^{-1}_1\tens \F^{-1}_2$ is the 
inverse\footnote{it exists as a formal power series because of 
(\ref{cond2bis})} of $\F$, then  
\berr
\p (m_q)    &=& m \circ(\p\tens\p),                      \label{defm}\\
\eps_q      &=& \eps\circ \p,                             \label{defc}\\
\p(S_q(u))  &=& \g^{-1} S(\p(u))\g,                      \label{def1}\\
\p(S^{-1}_q(u))  &=&  \g' S(\p(u))\g'^{-1},              \label{def1'}\\
(\p\tens\p) \Delta_q(u)&=& \F\Delta(\p(u))\F^{-1},        \label{defd}\\
(\p\tens\p)\RR &=& \F_{21} q^{t\over 2}\F^{-1}.        \label{defR}
\err
for any $u \in U_q(g)$.
Here $t:=\Delta(C)-\one\tens C-C\tens \one$ is 
the canonical invariant element 
in $U(g)\tens U(g)$, $C$ is the quadratic Casimir, and
\berr
\g  &=& S(\F_1^{-1}) \F_2^{-1},\qquad\qquad \g' =\F_2 S \F_1, \nn\\
\g^{-1} &=& \F_1  S \F_2 = S \g', \qquad 
        \g'^{-1} = S(\F^{-1}_2) \F^{-1}_1 = S \g.
\label{gammas}
\err
Moreover, $\g^{-1} \g'$ is central in $U(g)[[h]]$.
The undeformed maps\footnote{we will suppress the multiplication maps 
from now on} $m,\eps,\Delta,S$ have
been linearly extended from $U(g)$ to $U(g)[[h]]$; notice that $S^2 = 1$.
$\F_{21}$ is obtained from $\F$ by flipping the tensor product. 
This kind of notation will be used throughout from now on.
Coassociativity of $\Delta_q$ follows from the fact that the  
(nontrivial) coassociator
\be
\phi:=[(\Delta\tens \id)\F^{-1}](\F^{-1}\tens \one)(\one \tens \F)
[(\id \tens \Delta)\F]
\label{defphi}
\ee
is $U(g)$--invariant, i.e. 
$$
[\phi,\Delta^{(2)}(u)]=0
$$
for $u \in U(g)$. Here $\Delta^{(2)}$ denotes the usual 2--fold coproduct.

In the present paper, we only consider finite--dimensional
representations, i.e. operator algebras rather than the abstract ones.
Then the formal parameter $q=e^h$ can be replaced by a real
number close to 1, and all statements in this section still hold 
since the power series will converge. One could then identify the algebras
$U(su(2))$ with $U_q(su(2))$ (but not as coalgebras!) via the 
the isomorphism $\varphi$. We will usually keep $\varphi$ explicit, however, 
in order to avoid confusions.

It turns out that the twist $\F$ is not determined uniquely, but there
is some residual ``gauge freedom'' \cite{drinfeld_quasi},
\be
\F \rightarrow \F T
\label{F_gauge}
\ee
with an arbitrary symmetric $T \in U(g)[[h]]^{\tens 2}$ which commutes with 
$\Delta(U(g))$ and satisfies 
(\ref{cond2}), (\ref{cond2bis}). The symmetry of $T$ guarantees that $\RR$ 
is unchanged, so that $\F$ remains a twist from $(U(g)[[h]],m,\eps,\Delta,S)$
to $(U_q(q),m_q,\eps_q,\Delta_q,S_q)$.
We will take advantage of this below.

While for the twist $\F$, little is known apart apart from 
its existence, one can show \cite{fiore_twist} using results of 
Kohno \cite{ko} and Drinfel'd \cite{drinfeld_quasi} 
that the twists can be chosen 
such that the following formula holds:
\be
\phi=\lim_{x_0,y_0\rightarrow 0^+}\left\{x_0^{-\frac h{2\pi i} t_{12}}
\vec{P}\exp\left[-\frac h{2\pi i}\int\limits^{1-y_0}_{x_0}dx
  \left({t_{12}\over x}+{t_{23}\over x-1}\right)\right] 
   y_0^{\frac h{2\pi i} t_{23}}\right\} 
=\one + o(h^2).
\label{phi-integral}
\ee
Here $\vec P$ denotes the path--ordered exponential. 
Such twists were called ``minimal'' by Fiore \cite{fiore_twist}, who showed 
that they satisfy the following remarkable relations:
\berr
\one &=& \F \Delta \F_{1}(1 \tens (S \F_2)\gamma) \label{F_id_1},\\
 &=& (1\tens S\F_2 {\g'}^{-1})\F (\Delta \F_1)   \label{F_id_2} \\
 &=& \F \Delta \F_{2}((S \F_1)\gamma'^{-1}\tens 1) \label{F_id_3}\\
 &=&(\gamma^{-1}S\F_1^{-1}\tens 1)\Delta\F_{2}^{-1}\F^{-1} \label{F_id_4}\\
 &=&  \Delta \F^{-1}_2 \F^{-1}(\g' S\F^{-1}_1\tens 1)\label{F_id_5}\\
 &=&  (1\tens \gamma' S\F^{-1}_2)\Delta \F^{-1}_{1}\F^{-1} \label{F_id_6}
\err 
All coproducts here are undeformed. 
Furthermore, we add the following observation: let $(V_i,\tr)$ be \reps of 
$U(g)$ and $I^{(3)}\in V_1\tens V_2\tens V_3$ be an invariant tensor, so that 
$u \tr I^{(3)} \equiv \Delta^{(2)}(u) \tr I^{(3)} = \eps(u) I^{(3)}$
for $u \in U(g)$. 
Then the (component--wise) action of $\phi$ on $I^{(3)}$ is trivial:
\be
\phi \tr I^{(3)} = I^{(3)}.
\label{phi_trivial}
\ee 
This follows from (\ref{phi-integral}): observe that 
$t_{12}$ commutes with $t_{23}$ in the exponent,
because e.g. $(\Delta(C)\tens\one)$ can be replaced by $\one\tens\one\tens C$ 
if acting on invariant tensors. Therefore the 
path--ordering becomes trivial, and (\ref{phi_trivial}) follows.

\paragraph{Star structure.}

Consider on $U(su(2))[[h]]$ the (antilinear) star structure
\be
H^* = H ,\quad {X^{\pm}}^* = X^{\mp},
\label{U_star_cpct_1}
\ee
with $h^* = h$, since $q$ is real.
It follows e.g. from its explicit form \cite{zac} that the algebra
map $\p$ is compatible with this star, 
$$
\p(u)^* = \p(u^*).
$$
It was shown in \cite{jurco} that using a suitable gauge transformation
(\ref{F_gauge}), it is possible to choose $\F$
such that it is unitary,
\be
(\ast \tens \ast) \F = \F^{-1}.
\label{F_real}
\ee
Moreover, it was stated in  \cite{fiore_twist} without proof
that the following stronger statement holds:

\begin{prop}
Using a suitable gauge transformation (\ref{F_gauge}), 
it is possible to choose a twist $\F$ which for $q \in \reals$
is both unitary and minimal, so that (\ref{F_real}) and 
(\ref{F_id_1}) to (\ref{F_id_6}) hold. 
\label{F_minimal}
\end{prop} 
Since this is essential for us, we provide a proof in Appendix A.

\sect{Twisted $U_q(g)$--covariant $\star$ --product algebras}
\label{sec:twisting}

Let $(\A,\cdot,\tr)$ be an associative $U(g)$--module algebra,
which means that there exists an action 
\berr
U(g) \times \A &\rightarrow& \A, \nn\\
  (u,a) &\mapsto& u \tr a   \nn
\err
which satisfies $u \tr (ab) = (u_{(1)} \tr a) (u_{(2)} \tr b)$
for $a, b \in \A$. Here $\Delta(u) = u_{(1)} \tens u_{(2)}$ denotes the 
undeformed coproduct. Using the map $\p$ (\ref{defphi}), 
we can then define an action of $U_q(g)$ on $\A$ by
\be
u \tr_q a:= \p(u) \tr a,
\ee
or $u \tr_q a_i = a_j \pi^j_i(\p(u))$ in matrix notation.
This does {\em not} define a $U_q(g)$--module algebra,
because the multiplication is not compatible with the coproduct of $U_q(g)$.
However, one can define a new multiplication on $\A$ as follows: 
\be
a \star b := (\F^{-1}_1\tr a) \cdot(\F^{-1}_2\tr b) 
            = \cdot(\F^{-1} \tr(a\tens b))
\label{star}
\ee 
for any $a,b \in \A$. 
It is well--known \cite{majid_book} that  $(\A, \star, \tr_q)$ 
is now a $U_q(g)$--module algebra: 
\berr
u \tr_q (a \star b) &=& 
      \varphi(u)\tr \((\F^{-1}_1\tr a)\cdot(\F^{-1}_2\tr b)\) \nn\\ 
 &=& \cdot \((\Delta(\varphi(u)) \F^{-1}) \tr a \tens b\) \nn\\
 &=& \cdot \( (\F^{-1}(\varphi\tens\varphi)\Delta_q(u)) \tr a \tens b\) \nn\\
 &=& \star \( \Delta_q(u) \tr_q a \tens b\) \nn
\err
for $u\in U_q(g)$. In general, this product
$\star$ is not associative, but it is {\em quasiassociative},
which means that
\be
(a\star b)\star c = (\tilde\phi_1\tr a)\star\((\tilde\phi_2\tr b)
         \star(\tilde\phi_3\tr c)\).
\label{left_assoc}
\ee
where 
\be
\tilde\phi := (\one \tens \F)
   [(\id \tens \Delta)\F ][(\Delta\tens \id)\F^{-1}](\F^{-1}\tens \one) = 
    U_{\F} \; \phi \; U_{\F}^{-1}
\label{defphitilde}
\ee
with
$$
U_{\F} = (\one \tens \F) [(\id \tens \Delta)\F]\;\; \in U(g)^{\tens 3},
$$
which satisfies
$$
[\tilde\phi, \Delta^{(2)}_q(u)] = 0
$$
for $u \in U_q(g)$.
All this follows immediately from the definitions.
Moreover, the following simple observation will be very useful:

\begin{lemma}
In the above situation, 
\be
(a\star b)\star c =  a\star (b\star c) 
\label{triv_assoc}
\ee
if one of the factors $a,b,c \in \A$ is invariant under $U(g)$.
If $(\A, \cdot)$ is commutative, then any element $S \in \A$  which is 
invariant under the action of $U(g)$, 
$u \tr S = \eps(u)\; S$, is central in  $(\A, \star, \tr_q)$
\label{inv_central}
\end{lemma}
Note that invariance of an element $a \in \A$ under 
$U(g)$ is the same as invariance under $U_q(g)$.
\begin{proof}
This follows immediately from (\ref{cond2}) together with the definition of 
$\tilde \phi$. To see the last statement,
assume that $S$ is invariant. Then 
\berr
S \star a &=&  (\F_1^{-1} \tr S)\cdot(\F_2^{-1} \tr a)\nn\\
           &=& \cdot\(((\eps\tens \one)\F^{-1}) \tr(S \tens a)\)  \nn\\
           &=& S \cdot a = a \cdot S \nn\\
           &=&  a \star S
\err
for any $a\in \A$.
\end{proof}

For actual computations, it is convenient to use a tensor notation
as follows: assume that the elements
$\{a_i\}$ of $\A$ form a \rep of $U(g)$.
Denoting $\tilde\phi^{rst}_{ijk} = \pi^r_i(\tilde\phi_1)$
$\pi^s_j(\tilde\phi_2)\pi^t_k(\tilde\phi_3)$, 
equation (\ref{left_assoc}) can be written as 
\berr
(a_i\star a_j)\star a_k &=& a_r \star(a_s\star a_t) \tilde\phi^{rst}_{ijk},
       \;\; \mbox{or} \nn\\ 
(a_1\star a_2)\star a_3 &=&  a_1 \star(a_2\star a_3) \; \tilde\phi_{123}.
\err
The last notation will always imply a matrix multiplication as above.

Conversely, given a $U_q(g)$--module algebra $(\A, \star, \tr_q)$,
one can twist it into a $U(g)$--module algebra $(\A,\cdot,\tr)$ 
by 
$$
a \cdot b := (\p^{-1}(\F^{(1)})\tr_q a) \star(\p^{-1}(\F^{(2)})\tr_q b) 
$$ 
where of course $u \tr a = \p^{-1}(u) \tr_q a$.
Now if $(\A, \star,\tr_q)$ was associative, then 
$(\A, \cdot,\tr)$ is quasiassociative,
$$
a\cdot (b\cdot c) = \phi \tr_q^{(3)} \((a\cdot b) \cdot c\) 
 :=\((\phi_1\tr_q a)\cdot(\phi_2\tr_q b)\)
   \cdot(\phi_3\tr_q c).
$$
Such a twist was used in \cite{ours_1} to obtain the associative
algebra of functions on the $q$--deformed fuzzy sphere from the 
quasi--associative algebra of functions on $D2$--branes in the $SU(2)$
WZW model found in \cite{ars}.

\paragraph{Commutation relations and $\RR$--matrices.}

These twisted algebras have a more intrinsic
characterization, which is much more practical. Consider 
a commutative $U(g)$--module algebra $(\A, \cdot, \tr)$, and the associated
twisted $U_q(g)$--module algebra $(\A, \star, \tr_q)$ as defined above.
Observe that the definition (\ref{star}) is equivalent to 
\berr
a \star b &=& (\F_1^{-1} \tr a) \cdot (\F_2^{-1} \tr b) = 
            (\F_2^{-1} \tr b)\cdot (\F_1^{-1} \tr a) \nn\\
    &=& \cdot\( (\F^{-1} \F \F^{-1}_{21}) \tr (b\tens a) \) \nn\\
    &=& (\TR_2 \tr_q b) \star (\TR_1\tr_q a)
\label{star_R_calc}
\err
where we define
\be
\TR := (\p^{-1}\tens\p^{-1}) \F_{21}\F^{-1} = \TR^{-1}_{21}.
\ee 
In a given representation, this can be written as
\be 
a_i \star a_j =  a_k \star a_l \; \TR_{ij}^{lk}, 
     \;\; \mbox{or}\;\;\;  a_1 \star a_2 = a_2 \star a_1 \; \TR_{12}
\label{aa_CR}
\ee
where
\be
\TR^{ji}_{kl} = (\pi_k^j \tens  \pi_l^i) (\TR).
\label{TR-matrix}
\ee
Now there is no more reference to the ``original'' $U(g)$--covariant 
algebra structure. $\TR$ does not satisfy the quantum Yang--Baxter
equation in general, which reflects the non--associativity of the 
$\star$ product. However it does satisfy
\berr
\TR \TR_{21} &=& \one,  \label{triang}\\
\TR_{(12),3} := (\Delta_q \tens 1) \TR  &=& 
                  \tp_{312} \TR_{13} \tp^{-1}_{132} \TR_{23}\tp_{123} \\
\TR_{1,(23)} := (1 \tens \Delta_q) \TR  &=&
              \tp^{-1}_{231} \TR_{13} \tp_{213} \TR_{12} \tp^{-1}_{123},
\label{quasitr}
\err
as can be verified easily. This means that we are working with
the quasitriangular quasi--Hopf algebra \cite{drinfeld_quasi} 
$(U_q(g),\Delta_q,\tp,\TR)$, which is obtained from the ordinary Hopf algebra 
$(U(g),\Delta,\one,\one)$ by the Drinfel'd twist $\F$. 
In practice, it is much easier to work with $\TR$ than with $\F$. 
For $q \in \R$, one can in fact write 
\be
\TR = \RR \; \sqrt{\RR_{21} \RR_{12}}^{-1},
\ee
where $\RR$ is the usual universal $R$ --matrix (\ref{defR}) 
of $U_q(g)$, which does satisfy the quantum Yang--Baxter equation.
The product $(\RR_{21} \RR_{12})$ could moreover be expressed in terms 
of the  Drinfel'd--Casimir 
\be
v = (S_q \RR_2) \RR_1 q^{-H},
\label{v}
\ee
which is central
in $U_q(g)$ and satisfies $\Del(v) = (\RR_{21} \RR_{12})^{-1} v\tens v$.
The square root is well--defined on all the \reps 
which we consider, since $q$ is real.

\paragraph{Twisted Heisenberg algebras.}

Consider the $U(g)$--module algebra 
$(\A_H, \cdot, \tr)$ with generators $a_i$ and $a^{\dagger}_j$ in
some given irreducible representation and commutation relations 
\berr
[a^{\dagger}_i, a^{\dagger}_j] &=& 0 = [a_i,a_j], \nn\\
\[a^{\dagger}_i, a_j\] &=& (g_c)_{ij} 
\err
where $(g_c)_{ij}$ is the (unique) invariant tensor
in the given representation of $U(g)$. 
We can twist $(\A_H, \cdot, \tr)$ as above, and
obtain the $U_q(g)$--module algebra $(\A_H, \star, \tr_q)$. 
The new commutation relations among the generators can be 
evaluated easily:
\berr
a_1 \star a_2 &=&  a_2 \star a_1 \;  \TR_{12} , \nn\\
a^{\dagger}_1 \star a^{\dagger}_2 &=& 
           a^{\dagger}_2 \star a^{\dagger}_1 \; \TR_{12}, \nn\\
a^{\dagger}_1 \star a_2 &=& g_{12} + \; a_2\star a^{\dagger}_1\;\TR_{12}.
\label{a_del_CR}
\err
Here 
\be
 {g}_{n m} = (g_c)_{r s} \; \pi^{r}_{n}(\F_1^{-1}) \pi^{s}_{m}(\F_2^{-1})
\label{g_twist}
\ee  
is the unique rank 2 tensor which is invariant under the action 
$\tr_q$ of $U_q(g)$. A similar relation holds for the invariant tensor 
with upper indices:
\be
{g}^{n m} = \pi_{r}^{n}(\F_1) \pi_{s}^{m}(\F_2)\; g_c^{r s},
\label{g_twist_up}
\ee  
which satisfies ${g}^{n m} {g}_{m l} = \d^n_l$. 
In particular, it follows that
\be
a\star a:=  a_1\star a_2 \; g^{12} =  a_1 \cdot a_2\; (g_c)^{12},
\label{twisted_square}
\ee
therefore the invariant bilinears remain undeformed. This
is independent of the algebra of the generators $a_i$.

It is sometimes convenient to use the $q$--deformed antisymmetrizer
\cite{fiore_schupp}
$$
P^-_{12} = \F_{12} (1-\d^{21}_{12}) \F^{-1}_{12} 
    = (\one - P \tilde \RR)_{12}
$$
acting on the tensor product of 2 identical representations, where 
$P$ is the flip operator. Then the commutation relations
(\ref{aa_CR}) can be written as
\be
a_1\star a_2\; P^-_{12} = 0.
\ee
For products of 3 generators, the following relations hold: 
\begin{lemma}
\berr
a_{1} \star (a_2\star a_3) &=&  (a_2 \star a_3) \star a_{ 1}\;\TR_{1,(23)},
                             \label{aaa_braid}\\
a^{\dagger}_1 \star (a_2\star a_3\; g^{23}) &=&
   2 a_1 + (a_2\star a_3\; g^{23})\star a^{\dagger}_1, \label{del_asquare}\\
a^{\dagger}_1 \star (a_2\star a_3\; P^-_{23}) &=&
   (a_2\star a_3\; P^-_{23}) \star a^{\dagger}_1\; \TR_{1,(23)}.
\label{aP-a_braid}
\err

\label{twist_qa_lemma}
\end{lemma}
The proof is in the Appendix.

\subsection{Integration}
\label{subsec:integral}

From now on we specialize to $g = su(2)$, even though much of the following
holds more generally.
Let  $\{a_i\}$ be a basis of the spin $K$ \rep of $U(su(2))$ with integer $K$,
and consider the (free) commutative algebra $\A$ generated by these variables.
Let $g_c^{ij}$ be the (real, symmetric)
invariant tensor, so that $a\cdot a :=a_i a_j g_c^{ij}$
is invariant under $U(su(2))$, and $g_c^{ij} g_c^{jk} = \d^{ik}$. 
We now impose on $\A$ the star structure
\be
a_i^* = g_c^{ij} a_j,
\label{a_class_star}
\ee
so that $\A$ can be interpreted as the algebra of  
complex--valued functions on $\R^{2K+1}$; in particular, $a \cdot a$ is 
real. Then the usual integral on $\R^{2K+1}$ defines a 
functional on (the subset of integrable 
functions in a suitable completion of) $\A$, which satisfies
\berr
\int d^{2K+1}\!a\; u \tr f &=& \eps(u) \int d^{2K+1}\!a\; f, \nn\\
\(\int d^{2K+1}\!a\; f\)^* &=& \int d^{2K+1}\!a\; f^*
\err
for $u \in U(su(2))$ and integrable $f \in \A$. 
More general invariant functionals on $\A$
can be defined as  
\be
\langle f \rangle := \int d^{2K+1}\!a\; \rho(a\cdot a)\; f 
\label{inv_functional}
\ee
for $f \in \A$, where $\rho$ is a suitable real weight function. 
They are invariant, real and positive:
\berr
\langle u \tr f\rangle &=& \eps(u) \langle f\rangle, \nn\\
\langle f\rangle ^* &=& \langle f^* \rangle \nn\\
\langle f^* f\rangle  &\geq& 0
\label{functional_props}
\err
for any $u \in U(su(2))$ and $f \in \A$. 
As usual, one can then define a Hilbert space of 
square--(weight--) integrable functions by
\be
\langle f, g \rangle := \langle f^* g\rangle = 
   \int d^{2K+1}\!a\; \rho(a\cdot a)\; f^* g.
\label{GNS}
\ee
Now consider the twisted $U_q(su(2))$--module algebra $(\A, \star, \tr_q)$
defined in the previous section. We want to find an integral on $\A$
which is invariant under the action $\tr_q$ of $U_q(su(2))$. Formally, this is 
very easy: since the {\em space} $\A$ is unchanged by the twisting, 
we can simply use the classical integral again, and verify invariance
$$
\int d^{2K+1}\!a\; u \tr_q f = \int d^{2K+1}\!a\; \p(u) \tr f
 = \eps(\varphi(u)) \int d^{2K+1}\!a\; f 
 = \eps_q(u) \int d^{2K+1}\!a\; f.
$$
Notice that the algebra structure of $\A$ does not enter here at all.
The compatibility 
with the reality structure will be discussed in the next section. 

Of course we have to restrict to certain classes of integrable functions.
However, this is not too hard in the cases of interest. 
Consider for example
the space of Gaussian functions, i.e. functions of the form 
$P(a_i) e^{-c (a\cdot a)}$ with suitable (polynomial, say) $P(a_i)$.
Using (\ref{twisted_square}), this is the same as 
the space of Gaussian functions in the sense of the star product,
$P_{\star}(a_i) e^{-c (a\star a)}$. This will
imply that all integrals occuring in perturbation theory are well--defined. 
Furthermore, one can obtain a twisted sphere by imposing 
the relation $a \star a = a\cdot a = R^2$. On this sphere, 
the integral is well--defined for any polynomial functions. 
The integral over the twisted $\R^{2K+1}$ 
can hence be calculated by first integrating over the sphere and then 
over the radius.
Finally, we point out the following obvious fact:
\be
\langle P(a) \rangle  = \langle P_0(a) \rangle 
\ee
where 
$P_0(a) \in \A$ is the singlet part of the decomposition 
of the polynomial $P(a)$ under the action $\tr_q$ of $U_q(su(2))$,
or equivalently under the action $\tr$ of $U(su(2))$.

\sect{An $U_q(su(2))$--covariant operator formalism}
\label{sec:operator-form}

In the previous section, we defined quasi--associative algebras
of functions on arbitrary \rep spaces of $U_q(su(2))$.
We will apply this to the coefficients of the fields on $\S_{q,N}$ later.
However, there is an alternative approach 
within the framework of ordinary operators and representations,
which is essentially equivalent for our purpose. We shall follow here 
closely the constructions in \cite{fiore_twist}.
It seems that both approaches have their own advantages, 
therefore we want to discuss them both.

We first recall the notion of the semidirect product (cross--product) 
algebra, which is useful here.
Let $(\A,\cdot,\tr)$ be an associative $U(su(2))$--module algebra.
Then $U(su(2)) \smash \A$ is the vector space $\A \tens U(su(2))$,
equipped with the structure of an associative algebra defined by
$ua = (u_{(1)}\tr a) u_{(2)}$.
Here $u_{(1)} \tens u_{(2)}$ is the undeformed coproduct of $U(su(2))$.

In the following, we shall be interested in \reps of $\A$ which 
have a ``vacuum'' vector $\rangle$ such that all elements can be 
written in the form $\A \rangle$, i.e. 
by acting with $\A$ on the vacuum vector. In particular, 
we will denote with $V_{\A}$ the free left $\A$--module $\A \rangle$  
which as a vector space is equal to $\A$. This will be called the 
the ``left vacuum representation'' (or left regular representation).
Now any\footnote{provided the kernel of the \rep
is invariant under $U(su(2))$, which we shall assume.}  
such \rep of $\A$ can naturally be viewed as a 
\rep of $U(su(2)) \smash \A$, if
one declares the vacuum vector $\rangle$ to be a singlet under $U(su(2))$,
$$
u \rangle = \eps(u) \rangle,
$$
and $u \tr (a\rangle) = (u \tr a) \rangle$. One can then verify the relations
of $U(su(2)) \smash \A$.

Inspired by \cite{fiore_twist}, we define
for any $a \in \A$ the element
\be
\hat a:= (\F^{-1}_1 \tr a) \F^{-1}_2 \; \in U(su(2)) \smash \A.
\label{a_hat}
\ee
Using the definition of the Drinfel'd twist, it is immediate 
to verify the following properties:
\berr
\hat a \rangle &=& a \rangle \nn\\
\hat a \hat b \rangle &=& (a \star b) \rangle
\err
where $(a\star b)$ is the twisted multiplication on $\A$ defined in 
(\ref{star}). More generally, 
\be 
\hat a_1 \hat a_2 .... \hat a_k \rangle = 
(a_1 \star(a_2 \star (.... a_{k-1} \star a_k) ...)) \rangle 
\label{hata_astar}
\ee
for any $a_i \in \A$. Hence the elements $\hat a$ realize the twisted
product (\ref{star}) on $\A$, with this particular bracketing. 
If $c \in \A$ is a singlet, or equivalently $[c, U(su(2))] = 0$
in $U(su(2)) \smash \A$, then
\be
\hat c = c.
\label{c_equation}
\ee
If in addition the algebra $\A$ is commutative, then 
$\hat c$ is central in  $U(su(2)) \smash \A$. 
Moreover, the new variables $\hat a_i$ are 
automatically covariant under the quantum group $U_q(su(2))$,
with the $q$--deformed coproduct: denoting 
$$
\hat u:= \p(u) \in U(su(2))
$$
for $u \in U_q(su(2))$, one easily verifies
\be
\hat u \hat a = \widehat{u_1 \tr_q a}\; \widehat{u_2},
\label{uhat_ahat}
\ee
where $u_1 \tens u_2$ denotes the $q$--deformed coproduct. 
In particular,
\berr
\hat u \hat a \rangle    &=& u \tr_q a \rangle  
  = \widehat{u \tr_q a}\rangle \nn\\
\hat u \hat a \hat b\rangle &=& 
    (\widehat{u_1 \tr_q a}) (\widehat{u_2 \tr_q b})\rangle. \nn
\err 
Therefore $U_q(su(2))$ acts correctly on the $\hat a$--variables in the
left vacuum representation. 
More explicitly, assume that $\A$ is generated (as an algebra) 
by generators $a_i$ transforming in the spin $K$ \rep $\pi$ of $U(su(2))$, 
so that $u a_i = a_j \pi^j_i(u_{(1)}) u_{(2)}$. Then (\ref{uhat_ahat}) 
becomes
\be
\hat u \hat a_i = \hat a_j \pi^j_i(\widehat u_{1}) \widehat u_{2}. 
\ee
In general, the generators $\hat a_i$ will not satisfy 
closed commutation relations, even if the $a_i$ do.
However if $[a_i, a_j] = 0$, then one can verify that 
(cp. \cite{fiore_twist})
\be
\hat a_i \hat a_j = \hat a_k \hat a_l \; \mathfrak{R}^{lk}_{ij}
\label{aa_hat_CR}
\ee
where 
\be
\mathfrak{R}^{ij}_{kl} = (\pi^i_k \tens \pi^j_l\tens id) 
      (\tilde \phi_{213}\; \tilde R_{12}\; \tilde \phi_{123}^{-1})
     \quad  \in U(su(2)).
\label{R_phi}
\ee
Again, this involves only the coassociator and the universal 
$\RR$--matrix. Such relations for field operators 
were already proposed in \cite{mack_schom} on general grounds;
here, they follow from the definition (\ref{a_hat}).
In the case of several variables, one finds 
\be
\hat a_i \hat b_j = \hat b_k \hat a_l \; \mathfrak{R}^{lk}_{ij}.
\ee
Indeed, no closed quadratic commutation relations for 
deformed spaces of function with generators $a_i$ in arbitrary \reps of 
$U_q(su(2))$ are known, which has has been a major obstacle for defining
QFT's on $q$--deformed spaces. 
In the present approach, the generators $\hat a_i$ satisfy quadratic 
commutation relations which close only in the bigger algebra
$U(su(2)) \smash \A$. In general, they are not easy to work with.
However some simplifications occur if we use minimal  
twists $\F$ as defined in Section \ref{sec:drinfeld_twist}, 
as was observed by Fiore \cite{fiore_twist}:

\begin{prop}
For minimal twists $\F$ as in (\ref{phi-integral}), 
the following relation holds:
\be
g^{ij}\hat a_i \hat a_j = g_c^{ij} a_i  a_j.
\label{quadratic_hat}
\ee
Here
\be
g^{ij} = \pi_{r}^{i}(\F_1) \pi_{s}^{j}(\F_2)\; g_c^{rs} =
         g_c^{il} \pi^j_l(\g'),
\label{g_twist_3}
\ee
where $\g'$ is defined in (\ref{gammas}).
In particular if $\A$ is abelian, this implies
that $g^{jk}\hat a_j \hat a_k$ is central in $U(su(2)) \smash \A$.
\label{a2_central_prop}
\end{prop}
We include a short proof in the appendix for convenience.
This will be very useful to define a quantized field theory.
From now on, we will always assume that the twists are minimal.

\paragraph{Derivatives.}

Let $\A$ be again the free commutative algebra with generators $a_i$
in the spin $K$ \rep of $U(su(2))$, and
consider the left vacuum \rep $V_{\A} = \A \rangle$ of $U(su(2)) \smash \A$.
Let $\del_i$ be the (classical) derivatives,
which act as usual on the functions in $\A$. 
They can be considered as operators acting on $V_{\A}$,
and as such they satisfy the relations of the classical Heisenberg algebra, 
$\del_i a_j = g^c_{ij} + a_j \del_i$. 
Now define
\be
\hat \del_i :=  (\F^{-1}_1 \tr \del_i) \F^{-1}_2,
\label{del_hat}
\ee
which is an operator acting on $V_{\A} = \A \rangle$; in particular, it
satisfies $\hat \del_i \rangle = 0$.
Then the following relations hold:

\begin{prop}
For minimal $\F$ as in (\ref{phi-integral}), the operators 
$\hat a_i$,  $\hat \del_j$ acting on the left vacuum representation satisfy
\berr
\hat \del_i (g^{jk}\hat a_j \hat a_k) &=&
       2 \hat a_i + (g^{jk}\hat a_j \hat a_k) \hat \del_i, \label{del_a2}\\
\hat \del_i \hat a_j &=& g_{ij} + \hat a_k \hat \del_l \; 
                \mathfrak{R}^{lk}_{ij}.                    \label{del_a}
\err
\label{derivatives-prop}
\end{prop}
The proof is given in the appendix; the second relation (\ref{del_a})
is again very close to a result (Proposition 6) in \cite{fiore_twist},
and it holds in fact in $U(su(2)) \smash \A$.
Of course, the brackets in (\ref{del_a2}) were
just inserted for better readability, 
unlike in Lemma \ref{twist_qa_lemma} where they were essential.
If we have algebras with several variables in the 
same representation, then for example
\be
\hat \del_{a_i} (g^{jk}\hat b_j \hat a_k) =
       \hat b_i + (g^{jk}\hat b_j \hat a_k) \hat \del_{a_i}
\label{del_mixed}
\ee
holds, in self--explanatory notation.

One advantage of this approach compared to the quasi--associative
formalism in the previous section
is that the concept of a star is clear,
induced from Hilbert space theory. This will be explained next.

\subsection{Reality structure.}
\label{subsec:star}

Even though the results of this section are more general, 
we assume for simplicity that 
$\A$ is the free commutative algebra generated by the elements
$\{a_i\}$ which transform in the spin $K$ \rep  of $U(su(2))$ with integer $K$,
i.e. the algebra of complex--valued functions on $\R^{2K+1}$
(or products thereof). Then the classical integral defines an
invariant positive functional on $\A$ which satisfies 
(\ref{functional_props}), and
$V_{\A}$ becomes a Hilbert space (\ref{GNS}) 
(after factoring out a null space if necessary). 
Hence we can calculate the operator adjoint of the generators
of this algebra. By construction, 
$$
a_i^* = g_c^{ij} a_j
$$
where $g_c^{ij}$ is the invariant tensor, normalized such that  
$g_c^{ij} g_c^{jk} = \d^{ik}$.
As discussed above, $V_{\A}$ is a \rep of the semidirect product 
$U(su(2)) \smash \A$, in particular it is a
unitary \rep of $U(su(2))$. Hence the star on the generators of $U(su(2))$ is
$$
H^* = H ,\quad {X^{\pm}}^* = X^{\mp}.
$$
Now one can simply calculate the star of the twisted variables 
$\hat a_i \in U(su(2)) \smash \A$. The result is as expected: 

\begin{prop}
If $\F$ is a minimal unitary twist
as in Proposition \ref{F_minimal}, then
the adjoint of the operator $\hat a_i$ acting on
the left vacuum representation $\A\rangle$ is 
\be
\hat a_i^* = g^{ij} \hat a_j.
\ee
\label{a_star_real_prop}
\end{prop}
This is proved in the appendix, and it was already found in \cite{fiore_twist}.
It is straightforward to extend these results to the case of 
several variables $a^{(K)}_i, b^{(L)}_j, ...$ in different representations,
using a common vacuum $\rangle$. The star structure is always of the form
(\ref{a_star_real_prop}). 

If $\A$ is the algebra of functions on $\R^{2K+1}$, we have seen above
that the left vacuum \rep
$\A\rangle$ is also a \rep of the Heisenberg algebra
$\A_H$ with generators  $a_i, \del_j$.
Again we can calculate the operator adjoints, and the result is
\berr
\del_i^* &=& - g_c^{ij} \del_j, \nn\\
\hat \del_i^* &=& - g^{ij} \hat\del_j. \nn
\err
Of course, all these statements are on a formal level,
ignoring operator--technical subtleties.

\subsection{Relation with the quasiassociative $\star$ --product}
\label{subsec:QO_relation}

Finally, we make a simple but useful observation, which provides
the connection of the operator approach in this section with the 
quasiassociative approach of Section \ref{sec:twisting}.
Observe first that an invariant (real, positive (\ref{functional_props}))
functional $\langle \; \rangle$ on $\A$ extends trivially as a
(real, positive)
functional on $U(su(2)) \smash \A$, by evaluating the generators of
$U(su(2))$ on the left (or right) of $\A$ with the counit. Now
for any tensor $I^{i_1 ... i_k}$ of $U_q(su(2))$, denote
$$
I(\hat a) := I^{i_1 ... i_k} \hat a_{i_1} ... \hat a_{i_k}  \qquad
\in U(su(2)) \smash \A,
$$
and
\be
I_{\star}(a) := I^{i_1 ... i_k} a_{i_1}\star 
( ...  \star (a_{i_{k-1}} \star a_{i_k}) ... ) \qquad
\in \A.
\label{I_star_tensor}
\ee
Then the following holds:

\begin{lemma}
\begin{itemize}
\item[1)] If $I = I^{i_1 ... i_k}$ is an invariant tensor of $U_q(su(2))$, 
then $I(\hat a)$ as defined above commutes with $u \in U_q(su(2))$,
\be
[u, I(\hat a)] = 0 \qquad \mbox{in}\;\;U(su(2)) \smash \A.
\ee
\item[2)]
Let $s\rangle \in \A\rangle$ be invariant, i.e.
$u\cdot s\rangle  = \eps_q(u) s\rangle$, and 
$I$, ..., $J$ be invariant tensors of $U_q(su(2))$. Then
$$
I(\hat a) ... J(\hat a)\; s \rangle 
  = I_{\star}(a) \star .... \star J_{\star}(a)\; s \rangle.
$$
\item[3)]
Let $I,J$ be invariant, and $P = P^{i_1 ... i_k}$ be an arbitrary tensor 
of $U_q(su(2))$. Denote with 
$P_0$ the trivial component of $P$ under the action of $U_q(su(2))$. 
Then for any invariant functional $\langle \; \rangle$ on $\A$,
\berr
\langle I(\hat a) ... J(\hat a)\; P(\hat a)\rangle
 &=& \langle I(\hat a) ... J(\hat a)\; P_0(\hat a)\rangle \nn\\
  &=& \langle I_{\star}(a) \star .... \star J_{\star}(a)
           (P_0)_{\star}(a)\rangle  \nn \\
  &=& \langle I_{\star}(a) \star .... \star J_{\star}(a) \star
           P_{\star}(a)\rangle
\label{IJ_stuff}
\err
Moreover if $\A$ is abelian, 
then the $I(\hat a)$, $J(\hat a)$ etc. can be considered as central
in an expression of this form, for example
$$
\langle I(\hat a) ... J(\hat a)\; P(\hat a_i)\rangle
  = \langle P(\hat a_i)\; I(\hat a) ... J(\hat a) \rangle
 = \langle P(\hat a_i)\; J(\hat a) ... I(\hat a) \rangle
$$
and so on.
\end{itemize}
\label{star_hat_lemma}
\end{lemma}
The proof follows easily from (\ref{uhat_ahat}), (\ref{hata_astar}) and 
Lemma \ref{inv_central}. The stars between the invariant 
polynomials $I_{\star}(a), ... , J_{\star}(a)$ 
are of course trivial, and no brackets are needed.

\sect{Twisted Euclidean QFT}
\label{sec:QFT}

These tools can now be applied to our problem of quantizing
fields on the $q$--deformed fuzzy sphere $\S_{q,N}$.  
Most of the discussion is not restricted to
this space, but it is on a much more rigorous level there
because the number of modes is finite.
We will present 2 approaches, the first based on twisted $\star$
--products as defined in Section \ref{sec:twisting}, 
and the second using an operator formalism as in Section 
\ref{sec:operator-form}. Both have their own merits which seem 
to justify presenting them both. Their equivalence will follow from
Lemma \ref{star_hat_lemma}.

First, we discuss some basic requirements for a quantum field theory 
on spaces with quantum group symmetry. 
Consider a scalar field, and expand it in its modes as
\be
\Psi(x) = \sum_{K,n} \psi_{K,n}(x)\;  a^{K,n}.
\label{psi_field}
\ee
Here the $\psi_{K,n}(x) \in \S_{q,N}$ are a basis of 
the spin $K$ \rep of $U_q(su(2))$,
\be
u \tr_q \psi_{K,n}(x) = \psi_{K,m}(x) \pi^{m}_{n}(u),
\label{psi_covar}
\ee
and the coefficients $a^{K,n}$ transform
in the dual (contragredient) representation of $\tilde U_q(su(2))$,
\be 
u \; \ttr_q a^{K,n} = \pi^{n}_{m}(\tilde S u) \; a^{K,m}.
\label{a_covar}
\ee  
It is important to distinguish the Hopf algebras
which act on the coefficients $a^{K,n}$ and on the functions $\psi_{K,n}(x)$,
respectively.
The Hopf algebra $\tilde U_q(su(2))$ is obtained from $U_q(su(2))$
by flipping the coproduct and using the opposite antipode $\tilde S = S^{-1}$.
In particular, the $\RR$--matrix and the invariant tensors are
also flipped: 
\be
\tilde g^K_{nm} = g^K_{mn}
\label{g_tilde}
\ee
where $\tilde g^K_{nm}$ is the invariant tensor of $\tilde U_q(su(2))$.
The reason for this will become clear soon.
Moreover, it is sometimes convenient to express the
contragredient generators in terms of ``ordinary'' ones,
\be 
a_{K,n} = \tilde g^K_{nm}  a^{K,m}.
\label{lower_index}
\ee 
Then $\tilde U_q(su(2))$ acts as 
$$
u\; \ttr_q a_{K,n} = a_{K,m} \;  \pi^{m}_{n}(u).
$$  
We assume that the coefficients $a^{K,n}$ generate 
some algebra $\A$. This is not necessarily the algebra of field operators, 
which in fact would not be appropriate in the Euclidean case even for $q=1$. 
Rather, $\A$ could be the algebra 
of coordinate functions on configuration space (space of modes)
for $q=1$, and an analog thereof for $q \neq 1$.
The fields $\Psi(x)$ can then be viewed as ``algebra--valued distributions''
in analogy to usual field theory, by defining
$$
\Psi[f]:= \int\limits_{\S_{q,N}} \Psi(x) f(x) \quad \in \A
$$
for $f(x) \in \S_{q,N}$. Then the covariance properties (\ref{psi_covar}) and 
(\ref{a_covar}) could be stated as 
\be
u\; \ttr_q \Psi[f] =  \Psi[u \tr_q f],
\label{Psi_covariance}
\ee
using the fact that $\int (u\tr_q f) g = \int f (S(u) \tr g)$.

Our goal is to define some kind of correlation functions
of the form 
\be
\langle \Psi[f_1] \Psi[f_2] ... \Psi[f_k] \rangle \quad \in \C
\label{correlation_function}
\ee
for any $f_1, ..., f_k \in \S_{q,N}$, in analogy to the undeformed case.
After ``Fourier transformation'' (\ref{psi_field}), this amounts to 
defining objects
\be
G^{K_1,n_1; K_2,n_2; ...;K_k, n_k} := 
   \langle a^{K_1,n_1} a^{K_2,n_2} ... a^{K_k,n_k}\rangle =: 
   \langle P(a) \rangle
\label{correlation_function_a}
\ee 
where $P(a)$ will denote some polynomial in the $a^{K,n}$ from now on,
perhaps by some kind of a ``path integral''
$ \langle P(a) \rangle = \frac 1{\NN} \int \D a\; e^{-S[\Psi]}  P(a)$.
We require that they should satisfy at least the following properties, 
to be made more precise later:
\begin{itemize}

\item[(1)] {\em Covariance:}
\be
\langle u\; \ttr_q P(a) \rangle = \eps_q(u) \; \langle P(a) \rangle,
\ee
which means that the $G^{K_1,n_1; K_2,n_2; ...;K_k, n_k}$ are invariant 
tensors of $\tilde U_q(su(2))$,

\item[(2)] {\em Hermiticity:} 
\be
\langle P(a) \rangle^* = \langle P^*(a) \rangle
\ee
for a suitable involution $*$ on $\A$,

\item[(3)] {\em Positivity:}
\be
\langle P(a)^* P(a)\rangle \geq 0,
\ee

\item[(4)] {\em Symmetry} 

under permutations of the fields, in a suitable sense discussed below.
\end{itemize}
This will be our heuristic ``working definition'' of a 
quantum group covariant Euclidean QFT.

In particular, the word ``symmetry'' in (4) needs some explanation.
The main purpose of a symmetrization axiom is that it
puts a restriction on the number of degrees 
of freedom in the model, which in the limit $q \rightarrow 1$
should agree with the undeformed case. More precisely, the amount of 
information contained in the correlation functions 
(\ref{correlation_function_a}) should be the same as for $q=1$, i.e. 
the Poincare series of $\A$ should be the same.
This means that the polynomials in the $a^{K,n}$
can be ordered as usual, i.e. they
satisfy some kind of ``Poincare--Birkhoff--Witt'' property.
This is what we mean with ``symmetry'' in (4). In more physical terms, it
implies the statistical properties of bosons\footnote{we do not 
consider fermions here.}.

However, it is far from trivial how to impose such a ``symmetry'' on tensors 
which are invariant under a quantum group.
Ordinary symmetry is certainly not consistent with 
covariance under a quantum group.
One might be tempted to replace ``symmetry'' by some kind of invariance 
under the braid group which is 
naturally associated to any quantum group. This group is generally
much bigger than the group of permutations, however, 
and such a requirement is qualitatively different and
leaves fewer degrees of freedom. The
properties (3) and (4) are indeed very nontrivial requirements
for a QFT with a quantum group spacetime symmetry,
and they are not satisfied in the proposals that have been given up to 
now, to the knowledge of the authors.

Covariance (1) suggests that the algebra $\A$ generated by the 
$a^{K,n}$ is a $U_q(su(2))$--module algebra.
This implies immediately that $\A$ cannot be commutative, 
because the coproduct of $U_q(su(2))$ is not cocommutative. 
The same conclusion can be reached by contemplating the meaning 
of invariance of an action $S[\Psi]$, which will be clarified below.
One could even say that a second quantization
is required by consistency. As a further guiding line, 
the above axioms (1) -- (4) should be verified easily
in a ``free'' field theory.
 
In general, there is no obvious candidate for an associative algebra
$\A$ satisfying all these requirements. 
We will construct a suitable quasiassociative algebra $\A$ 
as a star--deformation of the algebra of functions on configuration space
along the lines of Section \ref{sec:twisting}, 
which satisfies these requirements.
Our approach is rather general and should be applicable 
in a more general context, such as for
higher--dimensional theories. Quasiassociativity implies that the 
correlation functions (\ref{correlation_function}) 
make sense only after specifying 
the order in which the fields should be multiplied (by explicitly
putting brackets), however different ways of bracketing are always
related by a unitary transformation. Moreover, 
the correct number of degrees of freedom is guaranteed by construction. 
We will then define QFT's which satisfy the above
requirements using a path integral over the fields $\Psi(x)$, 
i.e. over the modes $a^{K,n}$. An associative approach will also be presented
in Section \ref{subsec:associative_quant}, 
which is essentially equivalent.

\subsection{Star product approach}
\label{subsec:quant_1}

The essential step is as follows.
Using the map $\p$ (\ref{phi}), the coefficients 
$a^{K,n}$ transform also under the spin $K$ representation
of $U(su(2))$, via $u\; \ttr a^{K,n} = \p^{-1}(u)\;\ttr_q  a^{K,n}$.
Hence we can consider the usual commutative 
algebra $\A^K$ of functions on $\R^{2K+1}$ generated
by the $a^{K,n}$, and view it as a left $U(su(2))$--module
algebra $(\A^K, \cdot, \ttr)$. As explained in the Section \ref{sec:twisting}, 
we can then obtain from it 
the left $\tilde U_q(su(2))$--module algebra $(\A^K, \star,\ttr_q)$, 
with multiplication $\star$ as defined in (\ref{star}). 
More generally, we consider the 
left $\tilde U_q(su(2))$--module algebra $(\A, \star,\ttr_q)$
where $\A = \tens\limits_{K=0}^N \A^K$.
Notice that
the twist $\tilde \F$ corresponding to the reversed coproduct must be 
used here, which is simply $\tilde \F_{12} = \F_{21}$.
The reality issues will be discussed in Section \ref{subsec:associative_quant}.

\paragraph{Invariant actions.}

Consider the following candidate for an invariant action,
\be
S_{int}[\Psi] = \int\limits_{\S_{q,N}} \Psi(x) \star (\Psi(x) \star \Psi(x)).
\label{S_int_3}
\ee
Assuming that the functions on 
$\S_{q,N}$ commute with the coefficients, $[x_i, a^{K,n}] = 0$, 
this can be written as
\berr
S_{int}[\Psi] &=& \int\limits_{\S_{q,N}} \psi_{K,n}(x)\; \psi_{K',m}(x)\; 
                \psi_{K'',l} (x) \; a^{K,n}\star (a^{K',m}\star a^{K'',l})\nn\\
   &=& I^{(3)}_{K,K',K'';\; n,m,l} \; a^{K,n}\star (a^{K',m}\star a^{K'',l})
       \quad \in \A.
\label{S_example}
\err
Here\footnote{note that the brackets are actually not necessary here because 
of (\ref{phi_trivial}). For higher--order terms they are essential, however.} 
$I^{(3)}_{K,K',K'';n,m,l} =$ 
$\int\limits_{\S_{q,N}} \psi_{K,n}\; \psi_{K',m}\; \psi_{K'',l}\;$
is by construction an invariant tensor of $U_q(su(2))$, 
\be
I^{(3)}_{K,K',K'';n,m,l} \; 
                 \pi^n_r(u_{1}) \pi^m_s(u_{2}) \pi^l_t(u_{3})
 = \eps_q(u) \; I^{(3)}_{K,K',K'';r,s,t}.
\label{invar_I}
\ee
We have omitted the labels on the various representations.
Hence $S_{int}[\Psi]$ is indeed an invariant element of $\A$:
\berr
u\; \ttr_q S_{int}[\Psi] &=&  I^{(3)}_{K,K',K'';\; n,m,l} \;
    \pi^{n}_r (\tilde S u_{\wtilde{1}}) 
    \pi^{m}_{s}(\tilde Su_{\wtilde{2}})
    \pi^{l}_t (\tilde Su_{\wtilde{3}})\; 
      a^{K,r}\star (a^{K',s}\star a^{K'',t}) \nn \\
 &=& \eps_q(u) \; S_{int}[\Psi] \nn
\err
using (\ref{invar_I}), 
where $u_{\wtilde{1}} \tens u_{\wtilde{2}}\tens u_{\wtilde{3}}$
is the 2--fold coproduct of $u\in \tilde U_q(su(2))$;
notice that the antipode reverses the coproduct. 
This is the reason for using $\tilde U_q(su(2))$. 

In general, our actions $S[\Psi]$ will be polynomials
in $\A$, and we shall only consider invariant actions, 
\be
u \;\ttr_q S[\Psi] = \eps_q(u)\;  S[\Psi] \qquad \in \A,
\label{invar_actions}
\ee
for $u \in \tilde U_q(su(2))$.
It is important to note that by Lemma \ref{inv_central}, 
the star product of any such invariant actions 
is commutative and associative, even though the full algebra of
the coefficients $(\A, \star)$ is not. 
Moreover we only consider actions which are obtained 
using an integral over $\S_{q,N}$ as in (\ref{S_int_3}), which 
we shall refer to as ``local''.

In particular, consider the quadratic action 
$$
S_{2}[\Psi] = \int\limits_{\S_{q,N}} \Psi(x) \star \Psi(x),
$$
which can be rewritten as
\berr
S_{2}[\Psi]  &=& \int\limits_{\S_{q,N}} \psi_{K,n}(x)\; \psi_{K,m}(x)\; 
               a^{K,n}\star a^{K,m}
   = \sum_{K=0}^N \;  g^K_{nm} \; a^{K,n}\star a^{K,m} \nn\\
   &=& \sum_{K=0}^N \;  \tilde g^K_{mn} \; a^{K,n}\star a^{K,m}.
\label{S_mass}
\err
Here we assumed that the
basis $\psi_{K,n}(x)$ is normalized such that
\be
\int\limits_{\S_{q,N}} \psi_{K,n}(x) \psi_{K',m}(x)
= \d_{K,K'} \; g^K_{n,m}.
\label{psi_normaliz}
\ee 
This action is of course invariant,
$
u \;\ttr_q S_{2}[\Psi] = \eps_q(u) \; S_{2}[\Psi].
$
Moreover, the invariant quadratic actions agree
precisely with the classical ones. Indeed, the most general invariant
quadratic action has the form
\berr
S_{free}[\Psi]  &=& \hf \sum_{K=0}^N D_K g^K_{nm} \; a^{K,n}\star a^{K,m}\nn\\
     &=& \hf \sum_{K=0}^N \; D_K (g_c^K)_{nm}\; a^{K,n}\cdot a^{K,m} 
\label{quadratic_action}
\err
using (\ref{twisted_square}), for some $D_K \in \C$. 
This will allow to derive Feynman rules from Gaussian integrals as usual.

\paragraph{Quantization: path integral.}

We will define the quantization by 
a (configuration space) path integral, i.e. some kind of integration
over the possible values of the coefficients $a^{K,m}$.
This integral should be invariant under $\tilde U_q(su(2))$.
Following Section \ref{subsec:integral}, we 
consider $\A^K$ as the vector space of complex--valued functions on 
$\R^{2K+1}$, and use the usual classical integral over $\R^{2K+1}$.
Recall that the algebra structure of $\A^K$ does not enter here at all.
The same approach was used in \cite{grosse} to define the quantization of the 
undeformed fuzzy sphere, and an analogous approach is usually taken on 
spaces with a star product \cite{SW}.
Notice that $K$ is an integer, since we do not consider fermionic fields here.
Explicitly, let $\int d^{2K+1}\!a^{K}\; f$ be the integral 
of an element $f \in \A^K$ over $\R^{2K+1}$. It is 
invariant under the action of $\tilde U_q(su(2))$ 
(or equivalently under $U(su(2))$) 
as discussed in Section \ref{subsec:integral}:
$$
\int d^{2K+1}\!a^K\; u \;\ttr_q f = \eps_q(u)\; \int d^{2K+1}\!a^K\; f.
$$
Now we define
$$
\int \D \Psi \; f[\Psi] := \int \prod_K d^{2K+1}\!a^{K}\; f[\Psi],
$$
where $f[\Psi] \in \A$ is any integrable function (in the usual sense)
of the variables $a^{K,m}$.
This will be our path integral, which is by construction 
invariant under the action $\ttr_q$ of $\tilde U_q(su(2))$.

Correlation functions can now be defined as 
functionals of ``bracketed polynomials'' 
$P_{\star}(a) = a^{K_1,n_1} \star (a^{K_2,n_2} \star (...\star a^{K_l,n_l}))$
in the  field coefficients by
\be
\langle P_{\star}(a)\rangle := \frac{\int  \D \Psi\; e^{-S[\Psi]} P_{\star}(a)}
                                    {\int  \D \Psi\; e^{-S[\Psi]}}.
\label{correlation}
\ee
This is natural, because all invariant actions $S[\Psi]$
commute with the generators $a^{K,n}$. 
Strictly speaking there should be a factor $\frac 1{\hbar}$ in front
of the action, which we shall omit. In fact there are now 3 different
``quantization'' parameters: $\hbar$ has the usual meaning, while 
$N$ and  $q-q^{-1}$ determines a quantization or deformation of space. 

Invariance of the action $S[\Psi] \in \A$ implies
that 
\be
\langle u\; \ttr_q P_{\star}(a) \rangle  = 
             \eps_q(u)\; \langle P_{\star}(a) \rangle,
\label{P_invariance}
\ee
and therefore
\be
\langle P_{\star}(a) \rangle  = \langle (P_0)_{\star}(a) \rangle
\label{P_P_0}
\ee
where $P_0$ is the singlet part of the polynomial $P$, as in
Lemma \ref{star_hat_lemma}.
These are the desired invariance properties, and they would not hold
if the $a^{K,n}$ were commuting variables.
By construction, the number of
independent modes of a polynomial $P_{\star}(a)$ with given degree
is the same as for $q=1$. One can in fact order them, using
quasiassociativity together with the commutation relations (\ref{aa_CR}) 
which of course also hold under the integral:
\be
\langle P_{\star}(a) \star((a_i \star a_j - a_k \star a_l \; \TR_{ij}^{lk}) 
\star Q_{\star}(a))\rangle  = 0,
\label{aa_CR_corr}
\ee
for any polynomials $P_{\star}(a),  Q_{\star}(a) \in \A$. This can also be 
verified using the perturbative formula (\ref{correlator_Z}) below.
Therefore the symmetry requirement (4) of Section \ref{sec:QFT} is satisfied.
Moreover, the following cyclic property holds:
\be
\langle a_i \star P_{\star}(a) \rangle = 
\langle  P_{\star}(a) \star a_k \rangle\;  \tilde D^k_i, 
  \qquad \tilde D^k_i = \tilde g^{kn} \tilde g_{in}
\ee
for any $P_{\star}(a)$. This follows using (\ref{P_P_0}) and
the well--known cyclic property of the $q$--deformed invariant tensor
$\tilde g_{ij}$. 

In general, the use of quasiassociative algebras for QFT
is less radical than one might think, and it is consistent with results of
\cite{ars} on boundary correlation functions in BCFT.
Before addressing the issue of reality, we develop some tools
to actually calculate such correlation functions in perturbation theory.

\paragraph{Currents and generating functionals.}

One can now introduce the usual tools of quantum field theory. 
We introduce (external) currents $J(x)$ by
\be
J(x) = \sum_{K,n}  \psi_{K,n}(x) \; j^{K,n},
\label{currents}
\ee
where the new generators $j^{K,n}$ are included into the 
$\tilde U_q(su(2))$--module algebra $\A$, again 
by the twisted product (\ref{star}).
We can then define a generating functional 
\be
Z[J] = \frac 1{\NN} \int  \D \Psi \; e^{-S[\Psi] + \int \Psi(x)\star J(x)},
\label{generating_Z}
\ee
which is an element of $\A$ but depends only on the current variables.
Here $\NN = \int  \D \Psi \; e^{-S[\Psi]}$. Note that
$$
\int \Psi(x) \star J(x) =  \int J(x) \star \Psi(x),
$$
which follows e.g. from (\ref{quadratic_action}).
Invariance of the functional integral implies that 
\be
u\; \ttr_q Z[J] =  \eps_q(u)\; Z[J]
\label{Z_invar}
\ee
for any $u \in \tilde U_q(su(2))$, 
provided the actions $S[\Psi]$ are invariant.

It is now useful to introduce derivatives $\dl_{(j)}^{K,n}$ 
similar to (\ref{a_del_CR}), which together with the currents form a 
twisted (quasiassociative)
Heisenberg algebra as explained in the previous section:
$$
\dl_{(j) n}^K \star j^{K'}_m =  \d_{K,K'}\;  \tilde g^K_{nm} +  
         \; j^{K'}_r \star \dl_{(j) s}^{K} \;\TR_{nm}^{sr}
$$
By a calculation analogous to (\ref{del_asquare}), it follows that
$$
\dl_{(j)}^{K,n} \Big(\int \Psi(x)\star J(x)\Big) = a^{K,n} + 
  \Big( \int \Psi(x)\star J(x) \Big)  \dl_{(j)}^{K,n}.
$$ 
Recall that it is not necessary to put a star if one of the 
factors is a singlet.

This is exactly what we need. We conclude immediately that 
$[\dl_{(j)}^{K,n}, \exp(\int \Psi\star J)] =$ 
$a^{K,n} \exp(\int \Psi\star J)$, and 
by an inductive argument it follows that
the correlation functions (\ref{correlation}) can be written as 
\be 
\langle P_{\star}(a)\rangle  = 
      {{}_{{}_{J=0}}}\langle  P_{\star}(\dl_{(j)})\; Z[J] \rangle_{\dl=0}.
\label{correlator_Z}
\ee
Here 
$ {{}_{{}_{J=0}}}\langle ...  \rangle_{\dl=0}$ 
means ordering the derivatives to the right of the currents
and {\em then} setting $J$ and $\dl_{(j)}$ to zero.
The substitution of derivatives into the bracketed polynomial $P_{\star}$ is 
well--defined, because the algebra of the generators $a$ is the same 
as the algebra of the derivatives $\dl_{(j)}$. 

The usual perturbative expansion can now be obtained easily. 
Consider a quadratic action of the form
$$
S_{free}[\Psi] = \int\limits_{\S_{q,N}} \hf \Psi(x) \star D \Psi(x),
$$
where $D$ is an invariant (e.g. differential) operator on $\S_{q,N}$, so that 
$ D \Psi(x) = \sum\psi_{K,n}(x)\; D_K a^{K,n}$
with $D_K \in \C$. It then follows as usual that 
\be
Z_{free}[J] := \frac 1{\NN_{free}}\; \int  \D \Psi \; e^{- S_{free}[\Psi] +
             \int \Psi(x)\star J(x)} = 
            e^{ \frac 12  \int J(x)\star D^{-1} J(x)}.
\label{Z_free}
\ee
This implies that after writing the full action in the form 
$S[\Psi] =  S_{free}[\Psi] +  S_{int}[\Psi]$, one has
\berr
Z[J] &=& \frac 1{\NN}\;
             \int  \D \Psi \; e^{- S_{int}[\Psi]} e^{- S_{free}[\Psi]
                   + \int \Psi(x)\star J(x)}  \nn\\
   &=& \frac 1{\NN'}\; e^{- S_{int}[\dl_{(j)}]}\;  Z_{free}[J] \rangle_{\dl=0}.
\label{Z_perturb}
\err
This is the starting point for a perturbative evaluation.
In the next section, we shall cast this into a form which is even more 
useful, and show that the ``vacuum diagrams''
cancel as usual.

\paragraph{Relation with the undeformed case.}

There is a conceptually simple relation of all the above models which are
invariant under $\tilde U_q(su(2))$ with models on the 
undeformed fuzzy sphere which are invariant under $U(su(2))$, at the 
expense of ``locality''.
First, note that the space of invariant actions (\ref{invar_actions})
is independent of $q$. More explicitly,
consider an interaction term of the form (\ref{S_example}).
If we write down explicitly the definition of the $\star$ product
of the $a^{K,n}$ variables, then it can be viewed as an interaction
term of $a^{K,n}$ variables with a tensor which is invariant under 
the {\em undeformed} $U(su(2))$, obtained from $I^{(3)}_{K,K',K'';\; n,m,l}$
by multiplication with representations of $\F$.
In the limit  $q = 1$, this $\F$ becomes trivial. 
In other words, the above actions can also be viewed
as actions on undeformed fuzzy sphere $S^2_{q=1,N}$, with interactions
which are ``nonlocal'' in the sense of  $S^2_{q=1,N}$, i.e.
they are given by traces of products of matrices only to the lowest 
order in $(q-1)$. Upon spelling out the $\star$ product in the 
correlation functions (\ref{correlation}) as well, they can be considered as
ordinary correlation functions of a slightly nonlocal field theory
on $S^2_{q=1,N}$, disguised by the transformation $\F$.

In this sense, $q$--deformation simply amounts to some kind of 
nonlocality of the interactions. 
A similar interpretation is well--known in the context of
field theories on spaces with a Moyal product \cite{SW}.
The important point is, however, that one can 
calculate the correlation functions for $q \neq 1$ {\em without}
using the twist $\F$ explicitly, using only $\hat R$ --matrices and the 
coassociators $\tilde \phi$, which are much easier to work with. This 
should make the $q$--deformed point of view useful.
It is also possible to generalize these results to other $q$--deformed spaces.

\subsection{Associative approach}
\label{subsec:associative_quant}

In order to establish the reality properties of the field theories introduced
above, it is easier to use an alternative formulation, using the results
of Section \ref{sec:operator-form}. The equivalence of the two 
formulations will follow from Section \ref{subsec:QO_relation}.
This will also allow to define field operators
for second--quantized models in 2+1 dimensions in Section \ref{subsec:2plus1}.

Consider the left vacuum \reps $V_{\A} = \A\rangle$ of $\A = \tens_K \A^K$
introduced in Section \ref{sec:operator-form}, and define the 
operators\footnote{they should not be considered as field operators.}
\be
\hat a^{K,n} = (\tilde \F^{-1}_1 \tr a^{K,n}) \tilde \F^{-1}_2 \; \;
          \in \; U(su(2)) \smash \A
\ee
acting on $\A\rangle$. We can then more or less repeat all the constructions
of the previous section with $a^{k,n}$ replaced by $\hat a^{K,n}$,
omitting the $\star$ product. The covariance property (\ref{Psi_covariance}) 
of the field
\be
\hat \Psi(x) = \sum_{K,n} \psi_{K,n}(x)\;  \hat a^{K,n}
\label{psi_field-op}
\ee
can now be written in the form
$$
 \Psi[u \tr_q f] = u_{\tilde 1} \Psi[f] \tilde Su_{\tilde 2}.
$$
Invariant actions can be obtained by contracting the $\hat a^{K,n}$ 
with invariant tensors of $\tilde U_q(su(2))$, and satisfy
$$
[u, S[\hat\Psi]] = 0
$$
for\footnote{recall that as algebra, there is no difference between
$U(su(2))$ and $\tilde U_q(su(2))$.}
$u \in \tilde U_q(su(2))$. For example, any actions of the form 
$$
S[\hat \Psi] = 
\int\limits_{\S_{q,N}} \hf \hat \Psi(x) D \hat \Psi(x) + 
    \lambda \hat \Psi(x)\hat \Psi(x) \hat \Psi(x)\; = 
  S_{free}[\hat \Psi] + S_{int}[\hat \Psi]\;  \quad 
          \in \;U(su(2)) \smash \A
$$
are invariant, where $D$ is defined as before. 
Using Proposition \ref{a2_central_prop}, the quadratic invariant actions 
again coincide with the undeformed ones. In general, higher--order 
actions are elements of $U(su(2)) \smash \A$ but not of $\A$. Nevertheless
as explained in Section \ref{subsec:QO_relation},
all such invariant actions $S[\hat \Psi]$ are in one--to--one correspondence
with invariant actions in the $\star$--product approach, 
with brackets as in (\ref{I_star_tensor}). This will be understood from now on.

Consider again the obvious (classical) functional 
$\int \prod_K d^{2K+1}\!a^K\;$ on $\A$ (or $V_{\A}$) as in the previous 
section,
and recall from Section \ref{subsec:QO_relation} that it extends trivially
to a functional on $U(su(2)) \smash \A$, by evaluating $U(su(2))$ with $\eps$.
We will denote this functional by 
$\int \D \hat \Psi$. Define correlation functions of polynomials 
in the $\hat a^{K,n}$ variables as 
\be
\langle P(\hat a)\rangle := 
   \frac{\int \D \hat \Psi e^{-S[\hat \Psi]} P(\hat a)}
        {\int \D \hat \Psi e^{-S[\hat \Psi]}}
       = \langle P_0(\hat a)\rangle.
\ee
Here $P_0$ is again the singlet part of the polynomial $P$. 
Then  Lemma \ref{star_hat_lemma} implies
\be
\langle P(\hat a)\rangle = \langle P_{\star}(a)\rangle,
\ee
always assuming that the actions $S[\hat \Psi]$ are invariant
under $U_q(su(2))$. This shows the equivalence with the approach of 
the previous section. Moreover, 
\be
\langle P(\hat a)\;\hat a_i \hat a_j \;Q(\hat a)\rangle 
    = \langle P(\hat a) \;\hat a_k \hat a_l \; 
       \mathfrak{R}^{lk}_{ij} \; Q(\hat a) \rangle,
\label{aa_hat_CR_corr}
\ee
follows from (\ref{aa_hat_CR}), or from (\ref{correlator_Z_op}) below
on the perturbative level.

\paragraph{Currents and generating functionals.}
We can again extend $\A$ by other variables such as currents
\be
\hat J(x) = \sum_{K,n}  \psi_{K,n}(x) \; \hat j^{K,n} 
            \qquad \in U(su(2)) \smash \A,    \label{currents_op}
\ee
and consider the generating functional
\be
Z[\hat J] = \frac 1{\NN}\; \int  \D \hat \Psi \; e^{-S[\hat\Psi] + 
             \int \hat \Psi(x)\hat J(x)} \rangle
\label{generating_Z_op}
\ee
with $Z[0] = 1$.
This is defined as the element of $\A \rangle \cong \A$ obtained
after integrating over the $a^K$--variables; 
the result depends on the currents only. 
The brace $\rangle$ indicates that the explicit $U(su(2))$
factors in $U(su(2)) \smash \A$ are evaluated by $\eps$.
Again, Lemma \ref{star_hat_lemma}
implies that $Z[\hat J]$ agrees precisely with 
the previous definition (\ref{generating_Z}).

As explained in Section \ref{sec:operator-form}, one can consider also the
twisted derivative operators $\hat \dl_{(j)}^{K,n}$, 
which act on $\A\rangle$. Using Proposition \ref{derivatives-prop},
we can derive essentially the same formulas as in the previous section,
omitting the star product. In particular, (\ref{del_mixed}) implies that
$$
\hat \dl_{(j)}^{K,n} \Big(\int \hat \Psi(x) \hat J(x)\Big) = \hat a^{K,n} + 
  \Big(\int \hat \Psi(x) \hat J(x) \Big)  \hat \dl_{(j)}^{K,n}. 
$$  
Since invariant elements of $\A$ are central as was pointed out 
below (\ref{c_equation}), we obtain as usual 
\berr
\langle P(\hat a)\rangle  &=& 
     {{}_{{}_{J=0}}}\langle P(\hat \dl_{(j)})\; Z[\hat J] \rangle_{\dl=0}  
                                              \label{correlator_Z_op} \\
Z[\hat J] &=& \frac 1{\NN}\;\int  \D \hat \Psi \; e^{-( S_{free}[\hat\Psi]
         + S_{int}[\hat\Psi]) + \int \hat \Psi(x)\hat J(x)}\rangle \; 
   = \frac 1{\NN'}\; e^{- S_{int}[\hat\dl_{(j)}]}\;  
                            Z_{free}[\hat J] \rangle_{\dl=0}
                                             \nn \\
Z_{free}[\hat J] &=& \frac 1{\NN_{free}}\; 
                 \int  \D \hat \Psi \; e^{- S_{free}[\hat\Psi] +
             \int \hat \Psi(x) \hat J(x)}\rangle \; = 
         e^{ \frac 12  \int \hat J(x) D^{-1} \hat J(x)}\rangle.
                                             \label{Z_free_op}
\err
Even though these formulas can be used to calculate correlators 
perturbatively, there is a form which is more convenient 
for such calculations. To derive it, observe that
(\ref{del_a2}) implies
\be
\hat \dl_{(j)}^{K,n}  \; e^{ \frac 12  \int \hat J(x) D^{-1} \hat J(x)}
   = e^{ \frac 12  \int \hat J(x) D^{-1} \hat J(x)} \;
     (D_K^{-1}\;  \hat j^{K,n} + \hat \dl_{(j)}^{K,n});
\ee
one can indeed verify that the algebra of 
\be
\hat b^{K,n} =  D_K^{-1}\; \hat j^{K,n} + \hat \dl_{(j)}^{K,n}
\label{b_def}
\ee
is the same as the algebra of $\hat a^{K,n}$. 
Therefore (\ref{correlator_Z_op}) can be rewritten as 
\berr
\langle P(\hat a)\rangle &=& \frac 1{\NN'}\;
 {{}_{{}_{J=0}}}\big\langle P(\hat \dl_{(j)})\; e^{- S_{int}[\hat\dl_{(j)}]}\; 
       e^{ \frac 12  \int \hat J(x) D^{-1} \hat J(x)}\big\rangle_{\dl=0} \nn\\
 &=& \frac 1{\NN'}\; 
    {{}_{{}_{J=0}}}\langle e^{ \frac 12  \int \hat J D^{-1} \hat J}
         P(\hat b)\; e^{- S_{int}[\hat b]}\rangle_{\dl=0}\nn\\
&=& \frac{{{}_{{}_{J=0}}}\langle P(\hat b)\; 
                e^{- S_{int}[\hat b]}\rangle_{\dl=0}}
       {{{}_{{}_{J=0}}}\langle e^{- S_{int}[\hat b]}\rangle_{\dl=0}}.
\label{J_del_Wick}
\err 
To evaluate this, one reinserts the definition (\ref{b_def}) 
of $\hat b$ as a sum of derivative operators $\hat \dl$ and current generators 
$\hat j$. Each $\hat \dl$ must be ``contracted'' with a $\hat j$ 
to the right of it using the commutation relations (\ref{del_a}),
which gives the inverse propagator $D_K^{-1}$,
and the result is the sum of all possible complete contractions. 
This is the analog of Wick's theorem. The contractions can be indicated 
as usual by pairing up the $\hat b$ variables with a line, before 
actually reordering them. Then each contribution can be 
reconstructed uniquely from a given complete contraction; this
could be stated in terms of Feynman rules. 

One can also show that the denominator exactly cancels the 
``vacuum bubbles'' in the numerator, as usual. 
Indeed, consider any given complete contraction of a term
$$
\hat b ... \hat b\; \frac 1{n!} (S_{int}[\hat b])^n.
$$
Mark the set of vertices which are connected (via a series of contractions)
to some of the explicit $\hat b$ generators on the left with blue, 
and the others with red. Then 2 neighboring red and blue vertices
can be interchanged keeping the given contractions, 
without changing the result. This is because only the homogeneous 
part of the commutation relations\footnote{associativity helps here.}
(\ref{del_a}) applies, and all
vertices are singlets (cp. Lemma \ref{star_hat_lemma}). 
Therefore the red vertices
can be moved to the right of the blue ones, and their contractions are
completely disentangled. Then the usual combinatorics yields
\be
\langle P(\hat a)\rangle = 
    {{}_{{}_{J=0}}}\langle P(\hat b)\; e^{- S_{int}[\hat b]}
     \rangle_{\dl=0,\;\mbox{\scriptsize no vac}}
\label{Wick_novac}
\ee
in self-explanatory notation. Of course this also holds in the 
quasiassociative version, but the derivation is perhaps less transparent.

In general, it is not easy to evaluate these expressions explicitly, 
because of the coassociators. However the lowest--order 
corrections $o(h)$ where $q=e^{h}$ are easy to obtain, using the fact 
that $\tilde \phi = \one + o(h^2)$ for minimal twists (\ref{phi-integral}).
If we write
$$
R_{12} = \one + h r_{12}\; + o(h^2),
$$
then 
$$
\tilde R_{12} = R_{12} \sqrt{R_{21} R_{12}}^{-1} 
              = \one + \frac h2 (r_{12} - r_{21}) \; + o(h^2),
$$
which allows to find the leading $o(h)$ corrections to the undeformed 
correlation functions explicitly.

\paragraph{Reality structure.}
One advantage of this formalism is that 
the reality structure is naturally induced from the Hilbert
space $V_{\A}$, as explained in Section \ref{subsec:star}.
Using  Proposition \ref{a_star_real_prop} and noting that the $a^{K,m}$ are in 
the contragredient \rep of $U(su(2))$, it follows that
\be
(\hat a^{K,n})^* = \tilde g^K_{nm} \hat a^{K,m}.
\label{a_reality}
\ee
We shall assume that all the actions are real, 
$$
S[\hat\Psi]^*  = S[\hat\Psi];
$$
this will be verified in the examples below.
Moreover, the classical integral defines a real functional 
on $U(su(2)) \smash \A$. Hence we conclude that
the correlation functions satisfy 
\be
\langle P(\hat a)\rangle^* = \langle P(\hat a)^*\rangle.
\ee
One can also show that
\be
\psi_{I,i}(x)^* = g^I_{ij} \psi_{I,j}(x),
\label{psi_reality}
\ee
where $g^I_{ij}$ is normalized such that $g^I_{ij} = (g^I)^{ij}$.
Therefore
\be
\hat \Psi(x)^* 
 = \hat \Psi(x),
\label{psi_hat_reality}
\ee
using (\ref{g_tilde}). This is useful to establish the reality of actions.
Of course, one could also consider complex scalar fields.
Finally, the correlation functions satisfy the positivity property
\be
\langle P(\hat a)^* P(\hat a)\rangle \geq 0,
\ee
provided the actions are real.
This is a simple consequence of the fact that $ P(\hat a)^* P(\hat a)$
is a positive operator acting on the left vacuum representation, together 
with the positivity of the functional integral.
It is one of the main merits of the present approach.

\sect{Examples}

\subsection{The free scalar field}

Consider the action
\be
S_{free}[\Psi] = -\int\limits_{\S_{q,N}} 
    \hf \hat \Psi(x) \star \Delta \Psi(x).  
\ee
Here the Laplacian was defined in \cite{ours_1} using a differential 
calculus as $\Delta = \ast_H d \ast_H d$, and 
satisfies\footnote{it is rescaled from the one in \cite{ours_1} 
so that its eigenvalues are independent of $N$.} 
$$
\Delta \psi_{K,n}(x)\; = 
     \frac 1{R^2}\; [K]_q [K+1]_q \;\psi_{K,n}(x) \equiv D_K \;\psi_{K,n}(x),
$$
where $[K]_q = \frac{q^K - q^{-K}}{q-q^{-1}}$.
The basis $\psi_{K,n}(x)$ is normalized as in (\ref{psi_normaliz}).
The action is real by (\ref{psi_hat_reality}), and can be rewritten as 
$$
S_{free}[\hat\Psi] = - \sum_{K,n} \hf D_K \;\tilde g^K_{nm}
               \hat a^{K,m} \hat a^{K,n}  
  = - \sum_{K,n} \hf D_K \; (\tilde g^K)^{mn} \hat a_{K,m} \hat a_{K,n}, 
$$
using (\ref{lower_index}).
As a first exercise, we calculate the 2--point functions. 
From (\ref{correlator_Z_op}) and (\ref{Z_free_op}), one finds
\berr
\langle \hat a^K_n \hat a^{K'}_{n'} \rangle &=& 
 {{}_{{}_{J=0}}}\langle \hat \dl_n^K \hat \dl_{n'}^{K'} 
           Z_{free}[\hat J]\rangle_{\dl=0} \nn\\
  &=&  {{}_{{}_{J=0}}}\langle\frac 12\;\hat \dl_{n}^K \hat \dl_{n'}^{K'} 
  \;(\sum (\tilde g^K)^{rs}\;\hat j^K_r D_K^{-1}\hat j^K_s)\rangle_{\dl=0}\nn\\
  &=& D_K^{-1}\;{}_{J=0}\langle\hat\dl_{n}^K  
        \hat j^{K'}_{n'} \rangle_{\dl=0}  
  \; = D_K^{-1}\; \d^{K K'} \; \tilde g^K_{n n'},  \nn
\err
where (\ref{del_a2}) was used in the last line.
This result is as expected, and it could also be obtained by using 
explicitly the definition of the twisted operators $\hat a^K$.

The calculation of the 4--point functions is more complicated, 
since it involves the coassociator. To simplify the notation, 
we consider the (most complicated) case where
all generators $a^K$ have the same spin $K$, which will be suppressed.
The result for the other cases can then be deduced easily. 
We also omit the prescriptions $(\dl=0)$ etc. 
Using first the associative formalism, (\ref{J_del_Wick}) yields
\berr
\langle \hat a_{n} \hat a_{m}\hat a_{k} \hat a_{l}  \rangle  
 &=& \langle(D^{-1} \hat j_n + \hat\dl_n)
     (D^{-1} \hat j_m + \hat\dl_m) 
    (D^{-1} \hat j_k + \hat \dl_k)
    (D^{-1} \hat j_l + \hat \dl_l)   \rangle \nn\\
 &=& \langle\hat\dl_n (D^{-1} \hat j_m + \hat\dl_m) 
    (D^{-1} \hat j_k + \hat\dl_k) D^{-1} \hat j_l  \rangle \nn\\
 &=& D^{-2}\;\langle\hat\dl_n \hat j_m  
     \tilde g_{kl} \; + \; \hat\dl_n \hat\dl_m
     \hat j_k \hat j_l  \rangle  \nn\\
 &=& D^{-2}\;\langle\tilde g_{nm} \tilde g_{kl} 
     + \hat\dl_n \hat\dl_m \hat j_k \hat j_l\rangle  \nn
\err
To evaluate this, consider
\berr
\langle\hat\dl_n \hat\dl_m \hat j_k \hat j_l \rangle 
&=& \langle\hat\dl_n (\tilde g_{mk} + \hat j_a \hat\dl_b\;
      \mathfrak{R}^{ba}_{mk}) \hat j_l \rangle \nn\\
&=& \tilde g_{mk} \tilde g_{nl}  + 
     \langle \hat\dl_n \hat j_a \hat\dl_b\;\hat j_s\pi^s_l
     (\mathfrak{R}^{ba}_{mk}) \rangle \nn\\
&=& \tilde g_{mk} \tilde g_{nl} 
    + \tilde g_{na} \tilde g_{bs}
    (\tilde\phi_{213} \tilde R_{12} \tilde\phi^{-1})_{mkl}^{bas} \nn
\err
Collecting the result, we recognize 
the structure of Wick contractions which are given by the invariant tensor
for neighboring indices, but involve the $\tilde R$--matrix and the 
coassociator $\tilde \phi$ for ``non--planar'' diagrams.

To illustrate the quasiassociative approach, we calculate the same 
4--point function using the $\star$ product. Then
\berr
\langle a_{n} \star (a_{m}\star(a_{k}\star a_{l})) \rangle 
  &=&  \langle \dl_n \star \big((D^{-1} j_m + \dl_m)\star 
    ((D^{-1} j_k + \dl_k)\star  D^{-1}  j_l)\big)\rangle \nn\\
  &=& D^{-2}\; \tilde g_{nm} \tilde g_{kl} 
    + D^{-2}\; \langle \dl_n \star (\dl_m \star
     (j_k\star  j_l)) \rangle  \nn
\err
using an obvious analog of (\ref{J_del_Wick}). Now
\berr
 \!\! \langle\dl_n \star (\dl_m \star (j_k\star j_l))\rangle \!\!\!
 &=& \langle\dl_n \star ((\dl_{m'} \star j_{k'})\star j_{l'})\rangle
     \;(\tilde\phi^{-1})^{m'k'l'}_{mkl} \nn\\
 &=& \tilde g_{nl'}\tilde g_{m'k'}  (\tilde\phi^{-1})^{m'k'l'}_{mkl} + 
    \langle\dl_n \star ((j_{m''} \star \dl_{k''} \tilde R^{k''m''}_{m'k'})
    \star j_{l'})\rangle\;(\tilde\phi^{-1})^{m'k'l'}_{mkl} \nn\\
 &=& \tilde g_{nl}\tilde g_{mk} +
     \langle\dl_n \star (j_{m'} \star (\dl_{k'} \star j_{l'}))\rangle\;
    (\tilde\phi_{213}\tilde R_{12}\tilde\phi^{-1})^{k'm'l'}_{mkl} \nn\\
 &=& \tilde g_{nl}\tilde g_{mk} +
     \tilde g_{n m'} \tilde g_{k'l'}\;
    (\tilde\phi_{213}\tilde R_{12}\tilde\phi^{-1})^{k'm'l'}_{mkl}, \nn
\err
in agreement with our previous calculation; here
the identity (\ref{gphi_id}) was used.
As pointed out before, the corrections to order $o(h)$ can now be obtained
easily.

\subsection{Remarks on $N \rightarrow \infty$ and $\phi^4$ theory.}

The above correlators for the free theory
are independent of $N$, as long as 
the spin of the modes is smaller than $N$. Therefore one can define 
the limit $N \rightarrow \infty$ in a straightforward way,
keeping $R$ constant. In this limit, the algebra of functions on
the $q$--deformed fuzzy sphere becomes
\be
\eps^{ij}_k x_i x_j = R\;(q-q^{-1})\; x_k, \qquad g^{ij} x_i x_j = R^2,
\ee
which defines $S^2_{q,N=\infty}$. It 
has a unique faithful (infinite--dimensional) 
Hilbert space representation \cite{podles}.

In an interacting theory, the existence of the limit $N \rightarrow \infty$
is of course a highly nontrivial question.
Consider for example the $\phi^4$ model, with action
$$
S[\Psi] = \int\limits_{\S_{q,N}} 
    \hf \hat \Psi(x) \Delta \hat \Psi(x) + \hf m^2 \hat\Psi(x)^2
   + \lambda \hat \Psi(x)^4 \; = S_{free} + S_{int}
$$
which is real, using (\ref{psi_hat_reality}).
We want to study the first--order corrections in $\lambda$ 
to the 2--point function
$\langle \hat a^K_i \hat a^{K}_{j} \rangle$ using 
(\ref{Wick_novac}):
$$
\langle \hat a^K_i \hat a^{K}_{j} \rangle = {{}_{{}_{J=0}}}\langle 
   \hat b^K_i\; \hat b^K_j\; \Big(1 - \lambda \int\limits_{\S_{q,N}} 
      \psi^{I,k}(x)\psi^{J,l}(x)\psi^{L,m}(x)\psi^{M,n}(x)\;
     \hat b^I_k \hat b^J_l \hat b^L_m \hat b^M_n\Big)
      \rangle_{\dl=0,\;\mbox{\scriptsize no vac}} .
$$
We only consider the ``leading'' planar tadpole diagram.
It is given by any contraction of the $\hat b^K_i$ and $\hat b^K_j$ with 
$\hat b$'s in the interaction term, which does not involve ``crossings''. 
All of these contributions are the same, hence we assume that 
$j$ is contracted with $k$ and $i$ with $l$. Then $\hat b^L_m $
is contracted with $\hat b^M_n$, which gives 
$D_L^{-1} \tilde g^L_{mn}\; \d^{LM}$. Now 
$\psi^{L,m}(x)\psi^{L,n}(x)\; \tilde g^L_{mn}\; \in \S_{q,N}$ 
is invariant under $U_q(su(2))$ and therefore proportional to the 
constant function. The numerical factor can be 
obtained from (\ref{psi_normaliz}):
$$
\int \psi^{L,m}(x)\psi^{L,n}(x)\; \tilde g^L_{mn} = 
\tilde g_L^{mn}\tilde g^L_{mn} = [2L+1]_q = {}_q \dim(V^L).
$$
Here $V^L$ denotes the spin $L$ \rep of $U_q(su(2))$. 
Using $\int 1 = 4\pi R^2$,
the contribution to $\langle \hat a^K_i \hat a^{K}_{j} \rangle$ is 
$$
\tilde g^K_{il} \tilde g^K_{jk}\; \l\int\psi^{K,k}(x)\psi^{K,l}(x)
       \sum_{L = 0}^N\;  D_L^{-1}\; \frac 1{4\pi R^2} [2L+1]_q = 
\tilde g^K_{ij}\; \frac{\l}{4\pi}
     \sum_{L=0}^N\; \frac{[2L+1]_q}{[L]_q [L+1]_q + m^2 R^2},
$$
up to combinatorial factors of order 1.
Unfortunately this diverges linearly in $N$ for $N \rightarrow \infty$,
whenever $q \neq 1$. This is worse that for $q=1$, where the divergence is
only logarithmic. This is in contrast to a result of \cite{oeckl}, 
which is however in the context of a different concept of (braided) 
quantum field theory which does not satisfy our requirements in 
Section \ref{sec:QFT}, and hence is not a ``smooth deformation'' of
ordinary QFT. 
The contributions from the ``non--planar'' tadpole diagrams 
are expected to be smaller, because the
coassociator $\tilde \phi$ as well as $\tilde R$ are unitary.
At least for scalar field theories, this behavior could be 
improved by choosing another Laplacian such as $\frac{v-v^{-1}}{q-q^{-1}}$
which has eigenvalues $[2L(L+1)]_q$, where $v$ is the Drinfel'd Casimir 
(\ref{v}). Then all diagrams are convergent 
as $N \rightarrow \infty$. 
Finally, the case $q$ being a root of unity is much more subtle, 
and we postpone it for future work.

\subsection{Gauge fields}

The quantization of gauge fields $\S_{q,N}$ is less clear at present,
and we will briefly indicate 2 possibilities.
Gauge fields were introduced in \cite{ours_1} as
one--forms $B \in \Omega^1_{q,N}$. 
Here $\Omega^1_{q,N}$ is the subspace of one--forms in the 
$U_q(su(2))$--module algebra 
of differential forms on 
$\S_{q,N}$. It turns out that there is a basis of 3 independent
one--forms $\theta^a$ which commute with all functions on $\S_{q,N}$. 
It is then natural to expand the gauge fields in that basis,
\be
B = \sum B_a \theta^a.
\label{B_expand}
\ee 
The fact that there are 3 independent one--forms means that 
one component is essentially radial and should be considered
as a scalar field on the sphere; however, it is impossible to find
a (covariant) calculus with ``tangential'' forms only. 
Therefore gauge theory on $\S_{q,N}$ as presented here 
is somewhat different from the conventional picture, but
may nevertheless be very interesting physically \cite{ars_2}.

Actions for gauge theories are expressions in $B$ which involve 
{\em no} explicit derivative terms. Examples are
\be
S_3 = \int B^3, \quad 
S_2 = \int B \ast_H B,  \quad S_4 = \int B^2 \ast_H B^2,
\label{B_actions}
\ee
where $\ast_H$ is the Hodge star operator. 
The curvature can be defined as $F = B^2 - \ast_H B$.
The meaning of the field $B$ becomes more obvious if it is written  
in the form 
\be
B = \Theta + A
\label{B_A_split}
\ee
where $\Theta \in \Omega^1_{q,N}$ is the ``Dirac--operator''. 
While $B$ and $\Theta$ become singular in the limit $N \rightarrow \infty$, 
$A$ remains well--defined. In these variables, a more standard form 
of the actions is recovered, including Yang--Mills 
\be 
S_{YM} :=  \int F \ast_H F =  \int (dA + A^2) \ast_H (dA + A^2)
\label{YM}
\ee
and Chern--Simons 
\be
S_{CS} := \frac 13 \int B^3 - \frac 12 \int B \ast_H B = - \mbox{const}
    + \frac 12  \int A dA + \frac 23 A^3
\label{S_CS}
\ee
terms. For further details we refer to \cite{ours_1}.

Even though these actions (in particular the prescription 
``no explicit derivatives'') are very convincing and have the correct
limit at $q=1$, the precise meaning of gauge invariance is not clear. 
In the case $q=1$, gauge transformations have the form
$B_a \rightarrow U^{-1} B_a U$ for any unitary
matrix $U$, and actions of the above type are invariant. 
For $q \neq 1$, the integral is a quantum trace which contains an explicit 
``weight factor''$q^{-H}$, breaking this symmetry. 
There is however another symmetry of the above actions where  
$U_q(su(2))$ acts on the gauge fields $B_a$ as \cite{ours_1}
\be
B_a \rightarrow u_1 B_a S u_2
\label{gaugetransform}
\ee
or equivalently $B \rightarrow u_1 B S u_2$. 
This can be interpreted as a gauge transformation, 
leaving the actions
invariant for any $u \in U_q(su(2))$ with $\eps_q(u) = 1$, and it is
distinct from the rotations of $B$. 
There is no obvious extension to a deformed $U(su(N))$ invariance, however.
There is yet another $\tilde U_q(su(2))$
symmetry, rotating the frames $\theta^a$ only, i.e. mixing the 
components $B_a$. The rotation of the field $B$ is rather complicated 
if expressed in terms of the $B_a$, however.

The significance of all these different symmetries is not clear,
and we are not able to preserve them simultaneously at the quantum level.
We will therefore indicate 
two possible quantization schemes, leaving different symmetries manifest.

\paragraph{1) Quantization respecting rotation--invariance.}

First, we want to preserve the $U_q(su(2))$ symmetry corresponding
to rotations of the one--forms $\Omega^1_{q,N}$, which is 
underlies their algebraic properties \cite{ours_1}.
We shall moreover impose the constraint 
$$
d \ast_H B = 0,
$$
which can be interpreted as gauge fixing. It is invariant under rotations,
and removes precisely the null--modes in the Yang--Mills and Chern--Simons 
terms. We expand the field $B$ into irreducible \reps under this
action of $U_q(su(2))$:
\be
B = \sum_{K,n;\a} \;  \Xi^{\a}_{K,n}(x)\; b_{\a}^{K,n}.
\label{B_irreps}
\ee
Here $\Xi^{\a}_{K,n}(x) \in \Omega^1_{q,N}$ are one--forms 
which are spin $K$ \rep of $U_q(su(2))$ (``vector spherical harmonics''). 
The multiplicity is now generically $2$
because of the constraint, labeled by $\a$. 

To quantize this, we can use the same methods as in Section \ref{sec:QFT}.
One can either define a $\star$ product of the coefficients
$b_{\a}^{K,n}$ as discussed there, or introduce the operators
$\hat b_{\a}^{K,n}$ acting on a left vacuum representation. 
Choosing the star product approach to be specific, one
can then define correlation functions as 
\be
\langle P_{\star}(b)\rangle = \frac 1{\NN}\int  \D B \; 
                      e^{-S[B]} P_{\star}(b)
\label{correlation_B}
\ee
where $\D B$ is the integral over all $b_{\a}^{K,n}$, 
write down generating functions etc.
This approach has the merit that the remarkable solution $B = \Theta$
of the equation $F = 0$ in \cite{ours_1} 
survives the quantization, because the corresponding mode 
is a singlet (so that $\hat b_{\a}^{0,0} = b_{\a}^{0,0}$ is undeformed). 
Incidentally, observe that the bracketings 
$\int (B B)\ast_H (B B)$ and $\int B (B\ast_H(BB))$ in the star--product 
approach are equivalent, because of (\ref{phi-integral}).

\paragraph{2) Quantization respecting ``gauge invariance''.}

First, notice that there is no need for gauge fixing before quantization
even for $q=1$, since the group of gauge transformations is compact.
To preserve the symmetry (\ref{gaugetransform}) as well as the rotation
of the $\theta^a$, we expand $B$ into irreducible \reps under these
2 symmetries $U_q(su(2))$ and $\tilde U_q(su(2))$:
\be
B = \sum_{K,n;a} \;\psi_{K,n}(x)\theta^a\; \b_{a}^{K,n}.
\label{B_irreps_2}
\ee
Now $\b_{a}^{K,n}$ is a spin $K$ \rep of $\tilde U_q(su(2))$ and a spin 1
\rep of $U_q(su(2))$. These are independent and commuting
symmetries, hence
the quantization will involve their respective Drinfel'd twists
$\tilde\F$ and $\F$. In the associative approach of Section \ref{sec:QFT}
we would then introduce
$$
\hat \b_{a}^{K,n} = \b_{a'}^{K,n'} \pi^{a'}_a(\F^{-1}_1) 
\pi^n_{n'}(\tilde S \tilde\F^{-1}_1)  \F^{-1}_2 \tilde\F^{-1}_2 \;  
            \in \; (\tilde U(su(2)) \tens U(su(2))) \smash \A.
$$
To avoid confusion, we have used an explicit matrix notation here.
The rest is formally as before, and will be omitted.
One drawback of this approach is that the above--mentioned
solution $B = \Theta$ is somewhat obscured now: the corresponding
mode is part of $\b_{a}^{1,n}$, but not easily identified. 
Moreover, ``overall'' rotation invariance is not manifest in this 
quantization.




\subsection{QFT in $2_q +1$ dimensions, Fock space}
\label{subsec:2plus1}

So far, we considered 2--dimensional $q$--deformed
Euclidean field theory. In this section,
we will add an extra (commutative) time and define a 2+1--dimensional 
scalar quantum field theory on $\S_{q,N}$ with manifest 
$\tilde U_q(su(2)) \times \reals$ symmetry, 
where $\reals$ corresponds to time 
translations. This will be done using an operator approach,
with $q$--deformed creation and anihilation operators acting on a 
Fock space. The purpose is mainly to elucidate the meaning of the 
Drinfel'd twists as ``dressing transformations''.

We consider real scalar field operators of the form
\be
\hat \Psi(x,t) = \sum_{K,n} \;  \psi^{K,n}(x)\;\hat a_{K,n}(t)
  \; + \; \psi^{K,n}(x)^* \; \hat a^+_{K,n}(t)
\label{psi_field_t-op}
\ee
where 
\be
a^{(+)}_{K,n}(t)= U^{-1}(t)\; a^{(+)}_{K,n}(0)\; U(t)
\ee
for some unitary time--evolution operator 
$U(t) = e^{-iH t/\hbar}$; we will again put $\hbar = 1$.
The Hamilton--operator $H$ acts on some Hilbert space $\H$.
We will assume that $H$ is invariant under rotations, 
$$
[H,u] = 0
$$
where $u \in U(su(2))$ is an operator acting on $\H$;
recall that as (operator) algebra, $\tilde U_q(su(2))$ 
is the same as $U(su(2))$.
Rather than attempting some kind of quantization 
procedure, we shall assume that 
\be
\hat a^{(+)}_{K,n}(t) = \tilde \F^{-1}_1 \tr a^{(+)}_{K,n}(t)\; 
   \tilde \F^{-1}_2 \;  =  \; a^{(+)}_{K,m}(t)\; \pi^m_n (\tilde \F^{-1}_1)\;
        \tilde \F^{-1}_2  
\label{a_t_hat}
\ee
as in (\ref{a_hat}), where $a^{(+)}_{K,n}= a^{(+)}_{K,n}(0)$ 
are ordinary creation--and anihilation 
operators generating a oscillator algebra $\A$, 
\berr
[a_{K,n}, a^+_{K',n'}] &=& \d_{K K'}\; (g_c)_{n n'}, \nn\\
\;[a_{K,n}, a_{K',n'}] &=& [a^+_{K,n}, a^+_{K',n'}] = 0 \nn
\err
and act on the usual Fock space\footnote{note that
this is the same as the ``left vacuum \rep'' of the subalgebra
generated by the $a^+_{K,n}$, in the notation of Section 
\ref{sec:operator-form}.}
\be
\H = \oplus\; (a^+_{K,n} ...\; a^+_{K',n'} |0\rangle).
\label{Fock_a}
\ee
$\H$ is in fact a \rep of $U(su(2)) \smash \A$, and 
the explicit $U(su(2))$--terms in (\ref{a_t_hat}) are now understood 
as operators acting on $\H$. 
Hence the $\hat a^{(+)}_{K,n}(t)$ are some kind of dressed 
creation--and anihilation operators, whose equal--time commutation 
relations follow from (\ref{aa_hat_CR}), (\ref{del_a}):
\berr
\hat a_{K,n}\; \hat a^+_{K',n'} &=& \d_{K,K'} \; g_{n n'} + 
     \hat a^+_{K',l'}\; \hat a_{K,l} \; \mathfrak{R}^{l l'}_{n n'}, \nn\\
\hat a^+_{K,n}\; \hat a^+_{K',n'} &=&
     \hat a^+_{K',l'}\; \hat a^+_{K,l} \; \mathfrak{R}^{l l'}_{n n'}, \nn\\
\hat a_{K,n}\; \hat a_{K',n'} &=&
     \hat a_{K',l'}\; \hat a_{K,l} \; \mathfrak{R}^{l l'}_{n n'}  \nn
\err
where $\hat a^{(+)}_{K,n} = \hat a^{(+)}_{K,n}(0)$. The Fock space 
(\ref{Fock_a}) can equivalently be written as
\be
\H = \oplus \; \hat a^{+K,n} ... \;\hat a^{+K',n'} |0\rangle.
\label{Fock_ahat}
\ee
Here the main point of our construction of a quantum group covariant 
field theory is most obvious, namely that a symmetrization postulate 
has been implemented which restricts the number of states in the Hilbert 
space as in the undeformed case. This is the meaning of the postulate (4) in 
the introductory discussion of Section \ref{sec:QFT}.
One could even exhibit a (trivial) action of the symmetric group $S_n$
on the $n$--particle space, using the unitary transformation induced
by the Drinfel'd twist $\F$, as in \cite{fiore_schupp}. Moreover,
using (\ref{psi_reality}) and an analog of (\ref{a_reality})
it follows that
$$
\hat \Psi(x,t) ^* = \hat \Psi(x,t).
$$
One can also derive the usual formulas for time--dependent 
perturbation theory, if we assume that the Hamilton operator has the form
$$
H = H_{free} + V
$$
where
\be
H_{free} = \sum_{K=0}^N \;  D_K\; (\tilde g^K)^{nm} \;
        \hat a^+_{K,n} \hat a_{K,m}
  = \sum_{K=0}^N \;  D_K\; (g_c^K)^{nm} \; a^+_{K,n} a_{K,m},
\ee
and $V$ may have the form 
$$
V = \int\limits_{\S_{q,N}} 
    \hat \Psi(x) \hat \Psi(x) ... \hat \Psi(x).
$$
Using (\ref{del_mixed}), one can see that 
$$
[H_{free}, \hat a^+_{K,l}] = D_K\;  \hat a^+_{K,l}
$$
and similarly for $\hat a_{K,l}$. 
Therefore the eigenvectors of $H_{free}$  have the form 
$\hat a^{+K,n} ... \;\hat a^{+K',n'} |0\rangle$ with 
eigenvalues $(D_K + ... + D_{K'}) \in \reals$, and if $V=0$, then
the time evolution is given as usual by
$$
\hat a^+_{K,n}(t)= e^{-i D_K t/\hbar}\; \hat a^+_{K,n}, \qquad
\hat a_{K,n}(t)= e^{i D_K t/\hbar}\; \hat a_{K,n}.
$$
One can then go to the interaction picture if $V \neq 0$ and
derive the usual formula involving time--ordered products.
However one must now keep the time--ordering explicit, and 
there seems to be no nice formula for contractions of time--ordered
products. We shall not pursue this any further here.


The main point here is that the above
definitions are entirely within the framework of ordinary
quantum mechanics, with a smooth limit $q \rightarrow 1$ where
the standard quantum field theory on the 
fuzzy sphere is recovered. Again, one could also consider the limit 
$N \rightarrow \infty$ while keeping $q$ constant. The existence 
of this limit is far from trivial.
Moreover there is nothing special about 
the space $\S_{q,N}$ as opposed to other, perhaps higher--dimensional
$q$--deformed spaces, except the technical simplifications because
of the finite number of modes. 
This shows that there is no obstacle in principle 
for studying deformations of quantum field theory on such spaces.

\paragraph{Acknowledgements:}
We would like to thank C.-S. Chu, G. Fiore and J. Pawelczyk 
for useful and stimulating discussions, and J. Wess for support
and hospitality at the Max--Planck Institut f\"ur Physik in M\"unchen.
H. S. and J. M. are grateful for useful visits at the 
Erwin--Schr\"odinger Institut in Vienna, and H. G and J. M. thank 
K. Sibold and E. Zeidler for hospitality at the University of 
Leipzig resp. the MPI in Leipzig. H. S. also thanks the DFG
for a fellowship, as well as D. Schiff for hospitality at the 
Laboratoire de Physique Th\'eorique in Orsay.

\sect{Appendix A: some proofs}

\paragraph{\em Proof of Proposition \ref{F_minimal}:}

Assume that $\F$ is minimal, so that  (\ref{phi-integral}) holds.
We must show that it can be chosen such that $\F$ is unitary as well.
Define
$$
A:=\F_{23}(1\tens \Delta)\F, \qquad B:= \F_{12} (\Delta\tens 1) \F,
$$
so that $\phi = B^{-1} A$. From (\ref{phi-integral}) it follows that
$(*\tens *\tens *)\phi = \phi^{-1}$, 
hence $A A^* = B B^*$, and more generally
$$
f(A A^*) = f(B B^*)
$$
for functions $f$ which are defined by a power series. This also implies that
$$
A f(A^* A) A^* = B f(B^* B) B^*
$$
for any such $f$, hence
$$
\phi f(A^* A) = f(B^* B) \phi^{*-1} = f(B^* B) \phi.
$$
In particular we can choose $f(x) = \sqrt{x}$ which makes sense
because of (\ref{cond2bis}), and obtain
\be
\sqrt{B^* B}^{-1} \phi \sqrt{A^*A} = \phi.
\label{BA_id}
\ee
On the other hand, the element $T:= ((* \tens *)\F) \; \F$ commutes with 
$\Delta(u)$ because $(* \tens *) \Delta_q(u) = \Delta_q(u^*)$,
and so does $\sqrt{T}$, which is well--defined in 
$U(su(2))[[h]]$ since $\F = 1 + o(h)$. Moreover, $T$ is symmetric,
noting that
\be
(* \tens *) (\F_{21} \F^{-1}) = \F \F_{21}^{-1}
\ee
which follows from the well-known relation 
$(* \tens *) \RR = \RR_{21}$ for $q \in \reals$.
Therefore $T$ is an admissible gauge transformation, and 
$\F':= \F \sqrt{T}^{-1}$ is easily seen to be
unitary (this argument is due to \cite{jurco}).
In particular, since $\F^* \F$ commutes with $\Delta(u)$, it follows that
$A^* A = (\F_{23}^* \F_{23})(1\tens \Delta)(\F^* \F)$
and  $B^* B = (\F_{12}^* \F_{12})(\Delta\tens 1)(\F^* \F)$. 
Looking at the definition (\ref{defphi}), this means that
the left--hand side of (\ref{BA_id}) is the gauge transformation of $\phi$ 
under a gauge transformation
$\F \rightarrow \F':= \F \sqrt{T}^{-1}$, which makes $\F$ unitary. 
Therefore the coassociator is unchanged under this gauge transformation, 
hence it remains minimal.

\paragraph{\em Proof of Lemma \ref{twist_qa_lemma}:}

We simply calculate
\berr
a^{\dagger}_1 \star (a_2\star a_3)  &=&  
        (a^{\dagger}_1 \star a_2) \star a_3 \; \tilde\phi_{123}^{-1}\nn\\
  &=& (g_{12} + a_2 \star a^{\dagger}_1 \; \TR_{12}) 
              \star a_3\; \tilde\phi_{123}^{-1} \nn\\
  &=& g_{12}\; a_3 \;\tilde\phi_{123}^{-1} + 
        a_2 \star (a^{\dagger}_1 \star a_3)\;\tilde\phi_{213}\; 
        \TR_{12} \tilde\phi_{123}^{-1} \nn\\
 &=& g_{12}\; a_3 \;\tilde\phi_{123}^{-1} + 
        a_2 \star (g_{13} + \; a_3\star a^{\dagger}_1\;\TR_{13})\;
       \tilde\phi_{213}\; \TR_{12} \tilde\phi_{123}^{-1} \nn\\
&=& g_{12}\; a_3 \;\tilde\phi_{123}^{-1} +  
    a_2 g_{13} \tilde\phi_{213}\; \TR_{12} \tilde\phi_{123}^{-1} 
   +\;( a_2 \star a_3) \star a^{\dagger}_1\;\tilde\phi_{231}^{-1}\;
     \TR_{13}\; \tilde\phi_{213}\;  
     \TR_{12} \tilde\phi_{123}^{-1} \nn\\
 &=&  g_{12}\; a_3 \;\tilde\phi_{123}^{-1} 
      + a_2 \; g_{31} \tilde\phi_{231} \TR_{1,(23)} 
    + (a_2 \star a_3) \star a^{\dagger}_1\; \TR_{1,(23)}, \nn
\err
where (\ref{quasitr}) and $g_{31} \TR_{13} = g_{13}$ 
was used in the last step. Now
the first identity (\ref{aaa_braid}) follows immediately along these lines, 
omitting the inhomogeneous terms. 
To see the last one (\ref{aP-a_braid}), observe that
$$
g_{12} \tilde\phi_{123}^{-1} = (g_c)_{12} \F_{1,(23)}^{-1} \F_{23}^{-1} 
$$
because $g_{12} \F_{(12),3} = g_{12}$, and similarly
$$
g_{31} \tilde\phi_{231} \TR_{1,(23)} 
  =  (g_c)_{13}  \F_{1,(23)}^{-1} \F_{23}^{-1} .
$$
This implies that 
$$ 
\Big(g_{12}\; a_3 \;\tilde\phi_{123}^{-1} 
   + a_2\; g_{31} \tilde\phi_{231} \TR_{1,(23)}\Big) P^-_{23}
 = ((g_c)_{12} \; a_3+ (g_c)_{13}\; a_2) (1-\d^{23}_{32})
    \F_{1,(23)}^{-1}\F_{23}^{-1}  = 0
$$
where we used the fact that the undeformed coproduct is symmetric.
The second (\ref{del_asquare}) follows as above using  
$$
g_{31} \tilde\phi_{231} \TR_{1,(23)} g^{23} = \d_1^2,
$$
or simply from (\ref{twisted_square}).

\paragraph{\em Proof of Proposition \ref{a2_central_prop}:}

Relation (\ref{g_twist_3}) follows easily from 
\be
\pi^j_s(u) (g_c)^{rs} = \pi^r_l(Su) (g_c)^{lj}
\label{pi_S}
\ee
To prove (\ref{quadratic_hat}), consider
\berr
g^{ij}\hat a_i \hat a_j &=& g^{ij} \F_1^{-1} \tr a_i (\F_{2,1}^{-1}\F_a^{-1}) 
        \tr a_j \F_{2,2}^{-1} \F_b^{-1} \nn\\
  &=& a_k a_l g^{ij} \pi^k_i(\F^{-1}_1) \pi^l_j(\F_{2,1}^{-1}\F_a^{-1}) 
           \F_{2,2}^{-1} \F_b^{-1} \nn\\
  &=& a_k  a_l \pi^j_n(\g') (g_c)^{in} \pi^k_i(\F^{-1}_1)
         \pi^l_j(\F_{2,1}^{-1}\F_a^{-1}) \F_{2,2}^{-1} \F_b^{-1}. \nn
\err
Now we use
$\pi^k_i(\F^{-1}_1) (g_c)^{in} = (g_c)^{kr} \pi^n_r(S \F^{-1}_1)$, 
therefore
\berr
g^{ij}\hat a_i \hat a_j &=& a_k a_l (g_c)^{kr} 
 \pi^l_r(\F_{2,1}^{-1}\F_a^{-1}\g' S \F^{-1}_1)\F_{2,2}^{-1} \F_b^{-1} \nn\\
 &=& a_k a_l (g_c)^{kl}  \nn
\err
because of (\ref{F_id_5}).

\paragraph{\em Proof of Proposition \ref{derivatives-prop}:}

(\ref{del_a2}) follows easily from (\ref{quadratic_hat}):
\berr
\hat \del_i (g^{jk}\hat a_j \hat a_k)  &=&
       \hat\del_i ((g_c)^{jk} a_j a_k)  \nn\\
  &=& \del_n \pi^n_i(\F^{-1}_1) \F^{-1}_2  ((g_c)^{jk} a_j a_k)  \nn\\
  &=& \del_n \pi^n_i(\F^{-1}_1) ((g_c)^{jk} a_j a_k) \F^{-1}_2 \nn\\
  &=& 2 a_n \pi^n_i(\F^{-1}_1) \F^{-1}_2 
      +((g_c)^{jk} a_j a_k) \del_n \pi^n_i(\F^{-1}_1)\F^{-1}_2  \nn\\
  &=& 2 \hat a_i  
      +(g^{jk}\hat a_j \hat a_k) \hat \del_i, \nn
\err
as claimed. Next, consider
\berr
\hat \del_i \hat a_j &=& \del_n \pi^n_i(\F^{-1}_1) a_l 
    \pi^l_j(\F^{-1}_{2,1}\F^{-1}_a) \F^{-1}_{2,2} \F^{-1}_b \nn\\
    &=& (g_c)_{nl} \;\pi^n_i(\F^{-1}_1) 
         \pi^l_j(\F^{-1}_{2,1}\F^{-1}_a) \F^{-1}_{2,2} \F^{-1}_b 
    + a_l  \pi^l_j(\F^{-1}_{2,1}\F^{-1}_a)\del_n \pi^n_i(\F^{-1}_1)
             \F^{-1}_{2,2} \F^{-1}_b.   \nn
\err
The second term becomes 
$\hat a_k \hat \del_l \; \mathfrak{R}^{lk}_{ij}$ as in 
(\ref{aa_hat_CR}), and the first is
\berr
 (g_c)_{nl} \;\pi^n_i(\F^{-1}_1)
         \pi^l_j(\F^{-1}_{2,1}\F^{-1}_a) \F^{-1}_{2,2} \F^{-1}_b 
  &=& \pi^t_l(S\F^{-1}_1) (g_c)_{ti}\pi^l_j(\F^{-1}_{2,1}\F^{-1}_a) 
      \F^{-1}_{2,2} \F^{-1}_b  \nn\\
  &=&  (g_c)_{ti}\; \pi^t_j(S\F^{-1}_1 \F^{-1}_{2,1}\F^{-1}_a) 
      \F^{-1}_{2,2} \F^{-1}_b \nn\\
  &=&  (g_c)_{ti}\; \pi^t_j(\g) = 
      (g_c)_{ti}\; \pi^t_l(S \F_1^{-1}) \pi^l_j(\F_2^{-1}) \nn\\
  &=& (g_c)_{tl}\; \pi^t_i(\F_1^{-1})  \pi^l_j(\F_2^{-1}) = g_{ij} 
\label{gphi_id}
\err
using (\ref{F_id_4}).

\paragraph{\em Proof of Proposition \ref{a_star_real_prop}:}
Since $\pi$ is a unitary representation, we have
\berr
\hat a_i^* &=& \F_2 a_j^* \pi^i_j(\F_1) = 
             \F_2 a_k (g_c)^{kj}\pi^i_j(\F_1) \nn\\
  &=& \F_2 a_k (g_c)^{ni}\pi^k_n(S\F_1) \nn\\ 
  &=& a_l \pi^l_k(\F_{2,1}) (g_c)^{in} \pi^k_n(S\F_1)\F_{2,2} \nn\\
  &=& a_l \pi^l_k(\F_{2,1}S\F_1)\F_{2,2} g^{it} \pi^k_t(\g'^{-1})\nn\\
  &=& a_l \pi^l_t(\F_{2,1}S\F_1 \g'^{-1})\F_{2,2} g^{it} \nn\\
  &=& a_l \pi^l_t(\F^{-1}_1)\F^{-1}_2 g^{it} \nn\\
  &=& \hat a_t g^{it} \nn
\err
where (\ref{F_id_3}) was essential.

\end{document}